\documentclass[aps,pre,10pt,showpacs,reprint,floatfix,superscriptaddress]{revtex4}

\usepackage{times}
\usepackage{amsfonts}
\usepackage{amsmath}
\usepackage{amssymb}
\usepackage{graphicx}
\usepackage{epsfig}

\begin{document}

\title{A number-conserving linear response study of low-velocity ion stopping
in a collisional magnetized classical plasma}

\author{Hrachya B. Nersisyan}
\email{hrachya@irphe.am}
\affiliation{Institute of Radiophysics and Electronics, 0203 Ashtarak, Armenia}
\affiliation{Centre of Strong Fields Physics, Yerevan State University, Alex Manoogian
str. 1, 0025 Yerevan, Armenia}
\author{Claude Deutsch}
\affiliation{LPGP (UMR-CNRS 8578), Universit\'{e} Paris XI, 91405 Orsay, France}
\author{Amal K. Das}
\affiliation{Department of Physics, Dalhousie University, Halifax, Nova Scotia B3H 3J5, Canada}

\date{\today }

\begin{abstract}
The results of a theoretical investigation on the low-velocity stopping power of the ions moving in a
magnetized collisional plasma are presented. The stopping power for an ion is calculated employing linear
response theory using the dielectric function approach. The collisions, which leads to a damping of the
excitations in the plasma, is taken into account through a number-conserving relaxation time approximation
in the linear response function. In order to highlight the effects of collisions and magnetic field we
present a comparison of our analytical and numerical results obtained for a nonzero damping or magnetic
field with those for a vanishing damping or magnetic field. It is shown that the collisions remove the
anomalous friction obtained previously [Nersisyan \textit{et al}., Phys. Rev. E \textbf{61}, 7022 (2000)]
for the collisionless magnetized plasmas at low ion velocities. One of major objectives of this study is
to compare and contrast our theoretical results with those obtained through a novel diffusion formulation
based on Dufty-Berkovsky relation evaluated in magnetized one-component plasma models framed on target
ions and electrons.
\end{abstract}

\maketitle

\section{Introduction}
\label{sec:s1}

The energy loss of ion beams and the related processes in magnetized plasmas are important in many areas of physics
such as transport, heating and magnetic confinement of thermonuclear plasmas. The range of the related topics includes
ultracold plasmas \cite{kil07}, cold electron setups used for ion beam cooling \cite{sor83,ner07,mol03}, as well as
many very dense systems involved in magnetized target fusions \cite{cer00}, or inertial confinement fusion. This latter
thermonuclear scheme presently advocates a highly regarded fast ignition scenario \cite{tab94}, based on femtolaser
produced proton or heavier ion beams impinging a precompressed capsule containing a thermonuclear fuel \cite{rot01}
in it. Then, the magnetic field $B$ values up to $10^{10}$~G may be reached in the laboratory \cite{tat03}. Such a topic is also of intense
astrophysical concern \cite{win07}. These interaction geometries highlight low ion velocity slowing down (LIVSD) as
playing a fundamental role in asserting the confining capabilities and thermonuclear burn efficiency in dense and
strongly magnetized media.

For a theoretical description of the energy loss of ions in a plasma, there exist two standard approaches. The dielectric
linear response (LR) treatment considers the ion as a perturbation of the target plasma and the stopping is caused by the
polarization of the surrounding medium \cite{ner98,ner00,ner03,see98,wal00,ste01}. It is generally valid if the ion couples
weakly to the target. Alternatively, the stopping is calculated as the result of the energy transfers in successive binary
collisions (BCs) between the ion and the electrons \cite{ner09,zwi99,zwi02,toe02}. Here it is essential to consider
appropriate approximations for the screening of the Coulomb potential by the plasma \cite{ner07}.
However, significant gaps between these approaches involve the crucial ion stopping along magnetic field $\mathbf{B}$
and perpendicular to it. In particular, at high $B$ values, the BC predicts a vanishingly parallel energy loss, which remains
at variance with the nonzero LR one. Also challenging BC-LR discrepancies persist in the transverse direction, especially
for vanishingly small ion projectile velocity $v$ when the friction coefficient contains an anomalous term diverging
logarithmically at $v\to 0$ \cite{ner00,ner03}. In general when $v$ is smaller than target electron thermal velocity
$v_{e}$, the ratio $S(v)/v$, where $S(v)\equiv -dE/dx$ is the stopping power (SP), usually monitors a linear stopping
profile for highly ionized plasma with $B=0$ \cite{deu86} or $B\neq 0$ \cite{ner07}. An alternative approach, particularly
in the absence of any relevant experimental data, is to test various theoretical methods against comprehensive numerical
simulations \cite{ner07,ner09,zwi99,zwi02}. The latter exhibit high a level of numerical noise at large magnetic fields,
and in the $v\to 0$ limit, while keeping a plasma coupling below unity, which is precisely the domain of many important
applications of current interest.

With this background we report on a theoretical study of energy loss of a slow-velocity ion in a magnetized classical
plasma through a linear response approach which is constructed such that it conserves particle number. Broadly speaking,
there are two objectives of this paper. The first objective is to use the Bhatnagar-Krook-Gross (BGK) approach based on
the Boltzmann-Poisson equations for a collisional and magnetized classical plasma which is treated as a one-component
plasma (OCP). We use this approach to derive the dielectric function in a number-conserving manner and use this dielectric
function to calculate various aspects of energy loss of an ion moving in the plasma. This is done in Sec.~\ref{sec:s2}.
We would like to see how number-conservation and damping (due to collisions) affect the stopping power of an ion in
a low velocity range, i.e., for a slow ion. We should mention that for a colliisonal quantum plasma, e.g., a degenerate electron
gas (DEG) without magnetic field, Mermin \cite{mer70} and Das \cite{das75} considered the equation of motion for a single-particle
density matrix and derived the dielectric function in a number-conserving approach and in random phase approximation (RPA).
The main part of our first objective is to see to what extent the BGK approach can address the SP of a slow ion in a
magnetized one-component classical plasma. We will show that number conservation and collisions in such OCP have interesting
and experimentally observable effects on low-velocity SP. Now, one may expect various collision mechanisms in a magnetized
plasma. Because of their importance we have separately dealt with them in Sec.~\ref{sec:s3}. In Sec.~\ref{sec:s4} we present
detailed calculations of low-velocity SP in the BGK approach. Although the LR is normally used to calculate the SP of a fast
ion, we show that for the BGK model one can obtain useful and insightful results for low-velocity within the LR theory.

A second objective of our paper is to compare and contrast our theoretical results with those obtained through a different
method. The latter is specifically aimed at a low-velocity SP which is expressed in terms of velocity-velocity correlation
and hence to a diffusion coefficient. We refer to Dufty and Berkovsky (DB) \cite{duf95} for an exposition of this method.
The later part of Sec.~\ref{sec:s4} contains a brief account of the DB relation. Marchetti \emph{et al.} \cite{mar87} and
Cohen and Suttorp \cite{coh84} have calculated the relevant diffusion coefficients in a magnetized plasma through a hydrodynamical
and kinetic approaches, respectively. Recently Deutsch \emph{et al.} in Ref.~\cite{deu08} have suggested an alternative approach
for a calculation of low-velocity SP via the DB relation and by employing the diffusion coefficients for a magnetized OCP
\cite{mar87,coh84}, featuring either the target electrons or target ions. Since our main theoretical results are obtained in
a kinetic equation approach we have decided to devote Sec.~\ref{sec:s5} to an appraisal of these two approaches. In that
section we discuss results for an unmagnetized and magnetized plasma in the two approaches. This serves to highlight the
merits of the kinetic approach versus those in a hydrodynamical approach.

Section~\ref{sec:disc} contains some discussion and outlook. Appendix~\ref{sec:app1} can be consulted for an integral
representation of the dielectric function. The Coulomb logarithm where the dynamical polarization effects are neglected is
briefly discussed in Appendix~\ref{sec:app2}.

\section{Linear response formulation}
\label{sec:s2}

In this section we consider the main aspects of the linear response (LR) theory for the
ion-plasma interaction in the presence of an external magnetic field. Within the LR, the
electron plasma is described as a continuous, polarizable medium, which is represented by
the distribution function of the electrons $f(\mathbf{r},\mathbf{v},t)$. The evolution of
$f(\mathbf{r},\mathbf{v},t)$ is determined by the kinetic and Poisson equations. Usually
only a mean-field interaction between the electrons is considered and hard collisions are
neglected. This is valid for weakly coupled plasmas where the number of electrons in the
Debye sphere $N_{D}=4\pi n_{0e}{\lambda }_{e}^{3}=1/\epsilon \gg 1$ is very large. Here
$\epsilon$ is the plasma parameter, $n_{0e}$ and ${\lambda }_{e}=(k_{B}{T}_{e}/4\pi
n_{0e}e^{2})^{1/2}$ are the equilibrium density and the Debye length of electrons, respectively.

We consider a nonrelativistic projectile ion with charge $Ze$ and with a velocity $\mathbf{v}$,
which moves in a magnetized collisional and classical plasma at an angle $\vartheta $ with
respect to the constant magnetic field $\mathbf{B}$. We ignore any role of the electron
spin or magnetic moment due to the nonrelativistic motion of the ion and the plasma electrons.
We shall consider here the limit of heavy ions and neglect recoil effects. The strength of the
coupling between the moving ion and the electron plasma is given by the coupling parameter
\begin{equation}
\mathcal{Z}=\frac{Z/N_{D}}{(1+v^{2}/v_{e}^{2})^{3/2}} .
\label{eq:coup}
\end{equation}%
Here $v_{e}$ is the thermal velocity of the electrons. The derivation of Eq.~\eqref{eq:coup}
is discussed in detail in Ref.~\cite{zwic99}. The parameter $\mathcal{Z}$ characterizes
the ion-target coupling, where $\mathcal{Z}\ll 1$ corresponds to weak, almost linear coupling
and $\mathcal{Z}\gtrsim 1$ to strong, nonlinear coupling.

Let us now specify the kinetic equation for the collision-inclusive classical magnetized plasma.
The effect of collisions on the dielectric properties of the plasma is included, in a
number-conserving approximation, through a relaxation time $\tau =1/\gamma $, where $\gamma $ is
the collision frequency \cite{kra73}. For $\tau \to \infty $ the collision-inclusive kinetic
equation reduces to the collisionless Vlasov equation. Thus we consider the kinetic equation of the
collisional plasma within relaxation-time approximation (RTA) in which the collision term is of
the Bhatnagar-Gross-Krook(BGK)-type \cite{kra73},
\begin{equation}
\frac{\partial f}{\partial t}+\mathbf{v}\cdot \frac{\partial f}{\partial
\mathbf{r}}-\frac{e}{m_{e}}\left[ \mathbf{E}+\frac{1}{c}\left[ \mathbf{v}%
\times \mathbf{B}\right] \right] \cdot \frac{\partial f}{\partial \mathbf{v}}%
=-\gamma \left[ f-\frac{n_{e}}{n_{0e}}f_{0}\left( \mathbf{v}\right) \right] ,
\label{eq:1}
\end{equation}%
where the collision frequency $\gamma$ is a measure of damping of excitations in the plasma, and
\begin{equation}
n_{e}\left( \mathbf{r},t\right) =\int f\left( \mathbf{r},\mathbf{v},t\right)
d\mathbf{v} , \quad  n_{0e}=\int f_{0}\left( \mathbf{v}\right) d\mathbf{v}.
\label{eq:2}
\end{equation}%
Here $n_{e}(\mathbf{r},t)$ is the density of the electrons, $f_{0}(\mathbf{v})$ is the equilibrium distribution
function of the electrons in an unperturbed state. For instance, the distribution function $f_{0}(\mathbf{v})$
of the plasma electrons may be given by the Maxwell distribution. The right hand side of Eq.~\eqref{eq:1} is the
collision term in a relaxation-time approximation which was introduced by BGK in a number-conserving scheme.
It is easy to see that this form of collision term conserves the total number of particles. $\mathbf{E}$ is a
self-consistent electric field (see below) and $\mathbf{B}$ is treated as an external magnetic field. $\tau =1/\gamma$
is the relaxation time.

For a sufficiently small perturbations (LR treatment) we assume $f=f_{0}+f_{1}$, $n_{e}=n_{0e}+n_{1e}$, with
\begin{equation}
n_{1e}\left( \mathbf{r},t\right) =\int f_{1}\left( \mathbf{r},\mathbf{v}%
,t\right) d\mathbf{v} ,
\label{eq:3}
\end{equation}%
and the linearized kinetic equation becomes
\begin{equation}
\frac{\partial f_{1}}{\partial t}+\mathbf{v}\cdot \frac{\partial f_{1}}{%
\partial \mathbf{r}}-\Omega _{e}\left[ {\mathbf{v}}\times \mathbf{b}\right]
\cdot \frac{\partial f_{1}}{\partial \mathbf{v}}=\frac{e}{m_{e}}\mathbf{E}%
\cdot \frac{\partial f_{0}}{\partial \mathbf{v}}%
-\gamma \left( f_{1}-\frac{n_{1e}}{n_{0e}}f_{0} \right) .
\label{eq:4}
\end{equation}
Here $\mathbf{b}=\mathbf{e}_{z}=\mathbf{B}/B$ is the unit vector along the magnetic field,
$\Omega _{e}=eB/m_{e}c$ is the cyclotron frequency of the electrons, $\mathbf{E}=-\boldsymbol{\nabla}\varphi $,
$\varphi $ is the self-consistent electrostatic potential which is determined by the Poisson equation
\begin{equation}
\nabla ^{2}\varphi =-4\pi \rho _{0}\left( \mathbf{r},t\right) +4\pi e\int
f_{1}\left( \mathbf{r},\mathbf{v},t\right) d\mathbf{v} ,
\label{eq:5}
\end{equation}%
where $\rho_{0}$ is the density of the external charge.

We solve the system of equations \eqref{eq:4} and \eqref{eq:5} by space-time Fourier transforms. Because
of the cylindrical symmetry of the problem around the magnetic field direction $\mathbf{b}$, we introduce
cylindrical coordinates for the velocity $\mathbf{v}=\mathbf{e}_{x} v_{\bot }\cos\theta +\mathbf{e}_{y} v_{\bot}
\sin\theta +\mathbf{e}_{z}v_{\parallel}$ and the wave vector $\mathbf{k}=\mathbf{e}_{x} k_{\bot }\cos\psi +
\mathbf{e}_{y} k_{\bot} \sin\psi +\mathbf{e}_{z}k_{\parallel}$, where the symbols $\parallel $ and $\perp $
denote the components of the vectors parallel or perpendicular to the external magnetic field, respectively.
We next introduce the Fourier transforms of $f_{1}(\mathbf{r},\mathbf{v},t)$, $n_{1e}(\mathbf{r},t)$ and
$\varphi (\mathbf{r},t)$ with respect to variables $\mathbf{r}$ and $t$, $f_{1\mathbf{k},\omega}(\mathbf{v})$,
$n_{1e}(\mathbf{k},\omega)$ and $\varphi (\mathbf{k},\omega)$. Then the linearized kinetic equation \eqref{eq:4}
for the distribution function becomes
\begin{eqnarray}
&&\frac{\partial f_{1\mathbf{k}\omega}(\mathbf{v})}{\partial \theta}+\frac{i}{\Omega_{e}} \left(\mathbf{k}\cdot \mathbf{v}
-\omega -i\gamma \right)f_{1\mathbf{k}\omega} (\mathbf{v})  \label{eq:4a}  \\
&&=-\frac{ie}{m_{e}\Omega_{e}} \varphi \left(\mathbf{k},\omega\right)
\left(\mathbf{k} \cdot \frac{\partial f_{0}}{\partial \mathbf{v}}\right) + \frac{\gamma }{n_{0e}}n_{1e}\left(\mathbf{k},
\omega\right) f_{0}\left( \mathbf{v}\right) .    \nonumber
\end{eqnarray}
Assuming axially symmetric unperturbed distribution function, $f_{0}(\mathbf{v})=f_{0}(|v_{\parallel }|,v_{\perp })$,
Eq.~\eqref{eq:4a} can be formally integrated and the solution is
\begin{eqnarray}
&&f_{1\mathbf{k}\omega }\left( \mathbf{v}\right) =%
\frac{\gamma }{\Omega _{e} n_{0e}}f_{0}n_{1e}\left( \mathbf{k},\omega \right)
\int_{-\infty }^{\theta }\exp \left[ \frac{i}{\Omega _{e}}U\left( \theta
^{\prime }\right) \right] d\theta ^{\prime }  \label{eq:9} \\
&&-\frac{ie}{m_{e}\Omega _{e}}\varphi \left( \mathbf{k},\omega \right)
\int_{-\infty }^{\theta }\exp \left[ \frac{i}{\Omega _{e}}U\left( \theta
^{\prime }\right) \right] \left[ k_{\parallel }\frac{\partial f_{0}}{%
\partial v_{\parallel }}+k_{\perp }\cos \left( \theta ^{\prime }-\psi
\right) \frac{\partial f_{0}}{\partial v_{\perp }}\right] d\theta ^{\prime }, \nonumber
\end{eqnarray}%
where the lower limit of $\theta^{\prime}$-integration is chosen so as to take the integrand vanish. Here
\begin{equation}
U\left( \theta ^{\prime }\right) =\left( k_{\parallel }v_{\parallel }-\omega
-i\gamma \right) \left( \theta ^{\prime }-\theta \right) +k_{\perp }v_{\perp
}\left[ \sin \left( \theta ^{\prime }-\psi \right) -\sin \left( \theta -\psi
\right) \right] .
\label{eq:10}
\end{equation}%
The $\theta '$-integration in Eq.~\eqref{eq:9} can be performed using the Fourier series representation of the
exponential function \cite{gra80}. After straightforward integration we obtain
\begin{eqnarray}
&&f_{1\mathbf{k}\omega }\left( \mathbf{v}\right) =-e^{-iz_{e}\sin\left( \theta -\psi \right) }
\sum_{n=-\infty }^{\infty }\frac{e^{in\left( \theta -\psi \right)
}J_{n}\left(z_{e}\right)}{k_{\parallel }v_{\parallel }-\omega -i\gamma +n\Omega _{e}}  \label{eq:12} \\
&&\times \left[ \frac{i\gamma }{n_{0e}}f_{0}n_{1e}\left( \mathbf{k},\omega
\right) +\frac{e}{m_{e}}\varphi \left( \mathbf{k},\omega \right) \left(
k_{\parallel }\frac{\partial f_{0}}{\partial v_{\parallel }}+\frac{n\Omega
_{e}}{v_{\perp }}\frac{\partial f_{0}}{\partial v_{\perp }}\right) \right] , \nonumber
\end{eqnarray}
where $z_{e}=k_{\perp }v_{\perp }/\Omega_{e}$, $J_{n}$ is the Bessel function of the $n$th order. It
should be emphasized that Eq.~\eqref{eq:12} is a formal solution of the linearized kinetic equation
because the Fourier transformed electronic density $n_{1e}(\mathbf{k},\omega )$ remains unknown. We
now perform $\mathbf{v}$-integration in Eq.~\eqref{eq:12} and solve the obtained algebraic equation
with respect to the quantity $n_{1e}(\mathbf{k},\omega )$. Substituting this quantity into Fourier
transformed Poisson equation finally yields
\begin{equation}
\varphi \left( \mathbf{k},\omega \right) =\frac{4\pi \rho _{0}\left(\mathbf{%
k},\omega \right) }{k^{2}\varepsilon _{M}\left( \mathbf{k},\omega ,\gamma\right) } ,
\label{eq:17}
\end{equation}%
where $\varepsilon_{M}(\mathbf{k},\omega ,\gamma )$ is the collision-inclusive longitudinal dielectric
function of the magnetized plasma which is given by
\begin{equation}
\varepsilon _{M}\left( \mathbf{k},\omega ,\gamma \right) =1+\frac{\left(
\omega +i\gamma \right) \left[ \varepsilon \left( \mathbf{k},\omega ,\gamma
\right) -1\right] }{\omega +i\gamma Q\left( \mathbf{k},\omega ,\gamma \right) }
\label{eq:18}
\end{equation}%
with $\varepsilon (\mathbf{k},\omega ,\gamma )=\varepsilon _{e}(\mathbf{k},\omega +i\gamma )$,
$Q(\mathbf{k},\omega ,\gamma )=Q_{e}(\mathbf{k},\omega +i\gamma )$ and
\begin{eqnarray}
&&\varepsilon _{e}\left( \mathbf{k},\omega \right) =1-\frac{\omega _{e}^{2}%
}{k^{2}}\frac{2\pi }{n_{0e}}\int_{-\infty }^{\infty }dv_{\parallel
}\int_{0}^{\infty }v_{\perp }dv_{\perp }  \label{eq:19} \\
&&\times \sum_{n=-\infty }^{\infty }\frac{J_{n}^{2}\left(z_{e}\right)}
{k_{\parallel }v_{\parallel }+n\Omega _{e}-\omega -i0}%
\left( k_{\parallel }\frac{\partial f_{0}}{\partial v_{\parallel }}+\frac{%
n\Omega _{e}}{v_{\perp }}\frac{\partial f_{0}}{\partial v_{\perp }}\right) , \nonumber \\
&&Q_{e}\left( \mathbf{k},\omega \right) =\frac{2\pi }{n_{0e}}\int_{-\infty
}^{\infty }dv_{\parallel }\int_{0}^{\infty }f_{0}v_{\perp }dv_{\perp
}\sum_{n=-\infty }^{\infty }\frac{\left( k_{\parallel }v_{\parallel
}+n\Omega _{e}\right)J_{n}^{2}\left(z_{e}\right)}{k_{\parallel }v_{\parallel }+n\Omega _{e}-\omega -i0}. \label{eq:20}
\end{eqnarray}%
Here $\varepsilon _{e}(\mathbf{k},\omega )$ is the usual longitudinal dielectric function of the magnetized
collisionless and purely electron plasma (see, e.g., Ref.~\cite{kra73}) and $\omega _{e}=(4\pi n_{0e}e^{2}/m_{e})^{1/2}$
is the plasma frequency of the electrons. Similarly, $Q_{e}(\mathbf{k},\omega)$ refers to the electron plasma.
The dielectric function $\varepsilon_{M}(\mathbf{k},\omega ,\gamma )$ given by Eqs.~\eqref{eq:18}-\eqref{eq:20} has
been obtained in the BGK approach which is number-conserving. Note the exact relation $Q_{e}(\mathbf{k},0)=1$ which
holds independently of the initial distribution $f_{0}$ if the latter is normalized to the unperturbed electronic
density $n_{0e}$, see the second relation in Eq.~\eqref{eq:2}.

It is well known \cite{kra73} that the usual relaxation-time approximation can be obtained from Eq.~\eqref{eq:1} if
the collision term is written as $-\gamma (f-f_{0})$ and is equivalent to replacing $\omega$ by $\omega +i\gamma$ in
the collisionless dielectric function $\varepsilon_{e}(\mathbf{k},\omega )$. This procedure is inadequate because it
does not conserve the local particle number and does not lead to the Drude behavior at long wavelengths ($\mathbf{k}\to
0$). This is remedied in the BGK approach. The $\mathbf{k}\to 0$ case of the number-conserving dielectric function is
also of interest. Noting that $Q_{e}(0,\omega )=0$, from Eq.~\eqref{eq:18} we find
\begin{equation}
\varepsilon _{D}\left( \omega \right) =\frac{k_{\perp }^{2}}{k^{2}}%
\varepsilon _{\perp }\left( \omega \right) +\frac{k_{\parallel }^{2}}{k^{2}}%
\varepsilon _{\parallel }\left( \omega \right)
\label{eq:d1}
\end{equation}%
with
\begin{equation}
\varepsilon _{\perp }\left( \omega \right) =1+\frac{\omega _{e}^{2}\left(
\omega +i\gamma \right) }{\omega \lbrack \Omega _{e}^{2}-\left( \omega
+i\gamma \right) ^{2}]}, \quad  \varepsilon _{\parallel }\left( \omega
\right) =1-\frac{\omega _{e}^{2}}{\omega \left( \omega +i\gamma \right) } .
\label{eq:d2}
\end{equation}
The above results are a generalization of the Drude dielectric function for magnetized plasmas. Equations~\eqref{eq:d1}
and \eqref{eq:d2} are known also as a ``cold" plasma approximation (see, e.g., Ref.~\cite{kra73}) and can be alternatively
obtained from Eqs.~\eqref{eq:18}-\eqref{eq:20} assuming initial distribution function $f_{0}(\mathbf{v})=n_{0e}\delta
(\mathbf{v})$. For a simplicity we shall call the expressions \eqref{eq:19} and \eqref{eq:20} as a Bessel-function
representation of the dielectric function. For many practical applications, however, it is important to represent the
dielectric function in an alternative but equivalent integral form, see Appendix~\ref{sec:app1} for details.

\begin{figure*}[tbp]
\includegraphics[width=80mm]{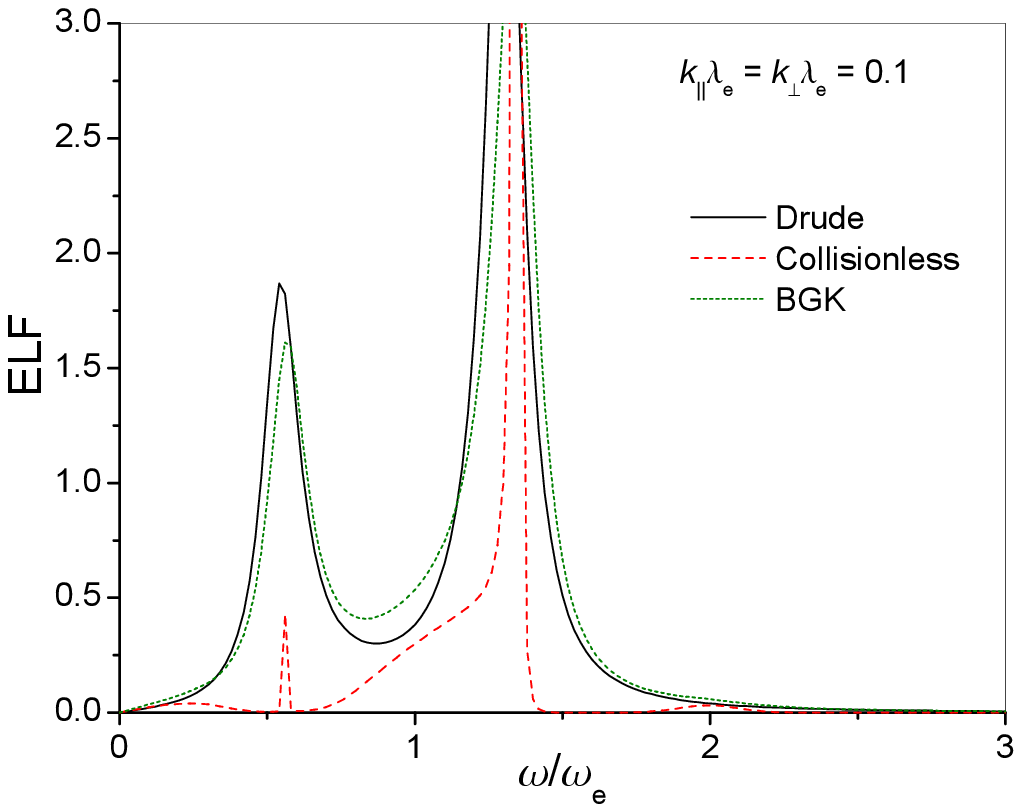}
\includegraphics[width=80mm]{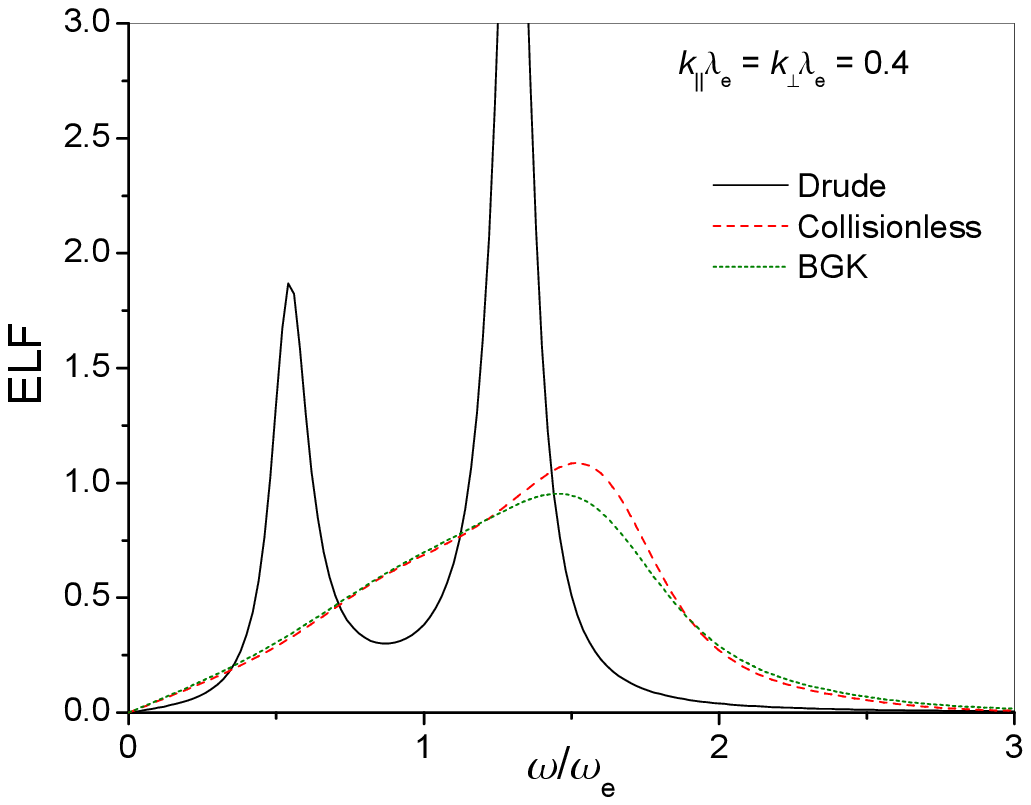}
\caption{(Color online) Generalized Drude (solid line), collisionless (dashed line) and number-conserving BGK
(dotted line) energy loss functions vs $\omega /\omega _{e}$ for $\Omega_{e}=\omega_{e}$, $\gamma =0.1\omega _{e}$
and $k_{\parallel }\lambda _{e}=k_{\bot }\lambda _{e}=0.1$ (left panel), $k_{\parallel}\lambda_{e}=k_{\bot}
\lambda _{e}=0.4$ (right panel).}
\label{fig:1}
\end{figure*}

Let us now specify the initial distribution function $f_{0}$ of the electrons. We consider the Maxwell
isotropic distribution function
\begin{equation}
f_{0}\left( v_{\parallel },v_{\perp }\right) =\frac{n_{0e}}{\left( 2\pi
\right) ^{3/2}v_{e}^{3}}\exp \left( -\frac{v_{\parallel }^{2}+v_{\perp }^{2}%
}{2v_{e}^{2}}\right) ,
\label{eq:22}
\end{equation}%
where $v_{e}=(k_{B}T_{e}/m_{e})^{1/2}$ is the thermal velocity of the electrons. The collision-inclusive
dielectric function then reads
\begin{equation}
\varepsilon \left( \mathbf{k},\omega ,\gamma \right) =1+\frac{1}{%
k^{2}\lambda _{e}^{2}}\left[ F_{1}\left( \mathbf{k},\omega \right)
+iF_{2}\left( \mathbf{k},\omega \right) \right] .
\label{eq:23}
\end{equation}%
Here
\begin{eqnarray}
&&F_{1}\left( \mathbf{k},\omega \right) =1+\sum_{n=-\infty }^{\infty }\frac{%
1}{\omega +n\Omega _{e}}\Lambda _{n}\left( \beta _{e}\right) \left[\omega
\mathcal{G}\left( x_{n},y\right) -\gamma \mathcal{F}\left( x_{n},y\right) \right] ,  \label{eq:24} \\
&&F_{2}\left( \mathbf{k},\omega \right) =\sum_{n=-\infty }^{\infty }\frac{1%
}{\omega +n\Omega _{e}}\Lambda _{n}\left( \beta _{e}\right) \left[ \omega
\mathcal{F}\left( x_{n},y\right) +\gamma \mathcal{G}\left( x_{n},y\right) \right] ,  \label{eq:25}
\end{eqnarray}%
are the real and imaginary parts of the generalized dispersion function of the collisional magnetized
plasma, respectively, and $x_{n}=(\omega +n\Omega _{e})/|k_{\parallel }|v_{e}$, $y=\gamma /|k_{\parallel }|v_{e}$,
$\beta _{e}=k_{\bot }^{2}a_{e}^{2}$, $a_{e}=v_{e}/\Omega _{e}$ is the cyclotron radius of the electrons,
$\Lambda _{n}(z)=e^{-z}I_{n}(z)$, $I_{n}(z)$ is the modified Bessel function of the $n$th order. Here we
have introduced the generalized Fried-Conte dispersion functions for the collisional plasma
\begin{eqnarray}
&&\mathcal{G}\left( x,y\right) =\frac{x}{\sqrt{2\pi }}\int_{-\infty }^{\infty
}\frac{\left( t-x\right) e^{-t^{2}/2}dt}{\left( t-x\right) ^{2}+y^{2}},
\label{eq:26} \\
&&\mathcal{F}\left( x,y\right) =\frac{xy}{\sqrt{2\pi }}\int_{-\infty
}^{\infty }\frac{e^{-t^{2}/2}dt}{\left( t-x\right) ^{2}+y^{2}}.
\label{eq:27}
\end{eqnarray}%
At vanishing $\gamma $ (at $y\to 0$) these functions become the usual Fried-Conte dispersion functions
\cite{fri61} of the collisionless plasma
\begin{eqnarray}
&&\mathcal{G}\left( x,0\right) =\frac{x}{\sqrt{2\pi
}}\int_{-\infty }^{\infty }\frac{e^{-t^{2}/2}dt}{t-x},  \label{eq:j1}   \\
&&\mathcal{F}\left( x,0\right) =\sqrt{\frac{\pi }{2}} xe^{-x^{2}/2} . \label{eq:j2}
\end{eqnarray}

The function $Q(\mathbf{k},\omega ,\gamma )$ which determines the dielectric function \eqref{eq:18}
is evaluated by inserting Eq.~\eqref{eq:22} into Eq.~\eqref{eq:20}. It is easy to see that for the
Maxwell isotropic distribution function \eqref{eq:22} the quantity $Q(\mathbf{k},\omega ,\gamma )$
is given by
\begin{equation}
Q\left( \mathbf{k},\omega ,\gamma \right) =\frac{\varepsilon _{e}\left( \mathbf{k},\omega +i\gamma \right) -1}{\varepsilon _{e}
\left( \mathbf{k},0\right) -1} =F_{1}\left( \mathbf{k},\omega \right) +iF_{2}\left( \mathbf{k},\omega \right) ,
\label{eq:21}
\end{equation}%
where
\begin{equation}
\varepsilon_{e}\left( \mathbf{k},0\right) =1+\frac{1}{k^{2}\lambda _{e}^{2}}
\label{eq:28}
\end{equation}%
is the static dielectric function which is not affected by the external magnetic field.

For ion stopping considerations, it is worth defining the energy loss function (ELF) $\mathrm{Im} [-1/\varepsilon
(\mathbf{k},\omega ,\gamma)]$. Figure \ref{fig:1} shows Drude, collisionless, and BGK energy loss functions
vs scaled frequency $\omega /\omega_{e}$ when $k_{\parallel}\lambda_{e}=k_{\bot}\lambda_{e}=0.1$ (left panel)
and $k_{\parallel}\lambda_{e}=k_{\bot}\lambda _{e}=0.4$ (right panel) for $\Omega_{e}=\omega_{e}$, $\gamma =
0.1\omega_{e}$. As has been mentioned above at small momentum $k\to 0$ the BGK energy loss function reproduces
the Drude energy loss function. And this is seen on the left panel of Fig.~\ref{fig:1}. Also at long wavelengths
(i.e., at small $k$) the BGK energy loss function is broadened due to the damping compared to the ELF with vanishing
damping.

The collision-inclusive dielectric function allows both physical insight and useful numerical estimates of the
influence of the collisions on energy loss. In an unmagnetized and degenerate electron gas the predicted effect
is a shorter lifetime and smaller mean free path of the plasmons resulting in considerable modifications of the
ELF \cite{ner08,ner02,ash80,abr94,bar06}. For the stopping of a single ion, the broadening of the plasmon peak with
increasing $\gamma$ shifts the threshold for the energy loss by plasmon excitation towards lower projectile
velocities. This increases the SP at low projectile velocities, compared to the collisionless result
\cite{ner08,ner02,ash80,abr94}. The situation with a present case of a classical and magnetized plasma including
the collisions may be quite different although collisional broadening of the ELF occurs also in this case.
This situation will be further discussed in Sec.~\ref{sec:s4}.

\section{Collision frequency in a magnetized plasma}
\label{sec:s3}

In Sec.~\ref{sec:s2} the effect of the collisions in a magnetized classical plasma has been introduced
in the dielectric function through a phenomenological but number-conserving collision term within the LR
theory. The model collision frequency $\gamma $ in solids and plasmas can be determined experimentally
or, alternatively can be calculated theoretically. For instance, in some investigations of ion stopping
in solid targets in the absence of a magnetic field, $\gamma $ was determined by fitting
$-\mathrm{Im} [\varepsilon ^{-1}(0,\omega ,\gamma )]$ to experimental optical energy loss functions (see, e.g.,
Refs.~\cite{ash80,abr94} and references therein). In addition the model relaxation time $\tau =1/\gamma $
can be estimated from the experimental data of the dc conductivity or the mobility in a plasma either with
or without external magnetic field. It should be emphasized that in general there are a number of physical
mechanisms which may contribute to the damping parameter $\gamma $. And contribution of each mechanism
depends strongly on the specific plasma conditions. We have not attempted here to evaluate the damping
parameter from first principles in the most general case but regard it rather as a model parameter.
In principle $\gamma $ can be calculated to varying degrees of approximations which may allow us to see
how the SP depends on the target properties and the magnetic field through their influence on $\gamma $.

In this section we briefly consider a fully ionized and a weakly coupled plasma where the contributions
of the Coulomb collisions to the frequency $\gamma$ may play a dominant role. This frequency, in our case, is determined
by electron-electron (e-e, $\gamma _{ee}$) and electron-ion (e-i, $\gamma _{ei}$) Coulomb collisions (if we do not
consider impurities). Thus, in contrast to Sec.~\ref{sec:s2}, we deal with a two-component electron-ion plasma (TCP)
accounting for the dynamics of plasma ions. The total effective frequency, in the limit of a weakly coupled plasma, can be approximated as
a sum of e-e and e-i collisions, $\gamma =\gamma _{ee}+ \gamma _{ei}$. In the absence of a magnetic field the theory of Coulomb collisions in a plasma
has been formulated by Spitzer \cite{spi62} (see also \cite{ram62}). In the last four decades or so the theory
has been further developed and extended. The recent book \cite{for06} summarizes the results obtained during
last four decades. However, to our knowledge, the relaxation processes in a magnetized plasma have not been
studied in as much detail as in an unmagnetized plasma, and only several theoretical attempts exist for this
case \cite{mon74,ich70,ich73,ich74,mat83,gli92,sil98,kor08} (see also the references therein). For a classical
plasma more complete expressions for the collision frequencies valid at arbitrary (but non-quantizing) magnetic
fields have been derived by Ichimaru \textit{et al.} and Matsuda \cite{ich70,ich73,ich74,mat83}, and by Montgomery
\textit{et al.} \cite{mon74} and by Silin \cite{sil98} with and without allowing for dynamical polarization effects
in plasma, respectively.

In Refs.~\cite{ich70,ich73} only e-i relaxation is considered. The generalization to the e-e case is straightforward.
The final result is summarized by a formula
\begin{equation}
\gamma _{e\alpha }=\frac{8\sqrt{2\pi }q_{\alpha }^{2}e^{4}n_{\alpha }\eta _{e\alpha }}{3m_{e}m_{\alpha }v_{e\alpha}^{3}}
\ln \Lambda _{e\alpha } ,
\label{eq:gamma1}
\end{equation}%
where $\alpha =e,i$ indicates the plasma species, $q_{e}=-1$, $q_{i}=Z_{i}$, $Z_{i}e$ is the charge
of plasma ion, $\eta _{ei}=1$, $\eta _{ee}=2^{1/2}$,
\begin{equation}
\ln \Lambda _{e\alpha }=\frac{1}{2\left( 2\pi \right) ^{3/2}\overline{v}%
_{e\alpha }^{3}}\int d\mathbf{k}\int_{-\infty }^{\infty }\frac{%
G_{e}\left( \mathbf{k},\omega \right) G_{\alpha }\left( \mathbf{k}%
,\omega \right) \omega ^{2}d\omega }{k_{\parallel }^{2}k^{4}\left\vert
\varepsilon_{ei}\left(\mathbf{k},\omega \right)\right\vert ^{2}},
\label{eq:gamma2}
\end{equation}
\begin{equation}
G_{\alpha }\left( \mathbf{k},\omega \right) =\sum_{n=-\infty }^{\infty
}\Lambda _{n}\left( k_{\perp }^{2}a_{\alpha }^{2}\right) \exp \left[ -\frac{%
\left( \omega -n\Omega _{\alpha }\right) ^{2}}{2k_{\parallel }^{2}v_{\alpha
}^{2}}\right] ,
\label{eq:gamma3}
\end{equation}%
$v_{\alpha }^{2}=k_{B}T_{\alpha }/m_{\alpha }$, $v_{e\alpha }^{2}=v_{e}^{2}+v_{\alpha}^{2}$,
$\overline{v}_{e\alpha }^{-2}=v_{e}^{-2}+v_{\alpha}^{-2}$,
$a_{\alpha }=v_{\alpha }/\Omega _{\alpha }$, $\Lambda _{n}(z)=e^{-z}I_{n}(z)$. Also $T_{i}$, $m_{i}$,
$v_{i}$ and $\Omega _{i}=Z_{i}eB/m_{i}c$ are the temperature, the mass, the thermal velocity and the
cyclotron frequency of the plasma ions ($a_{i}$ is the cyclotron radius), respectively. The quantity
$\ln\Lambda _{e\alpha}$ is the generalized Coulomb logarithm for a magnetized plasma. Here
$\varepsilon_{ei}(\mathbf{k},\omega )$ is the longitudinal dielectric function of a magnetized and
collisionless electron-ion TCP (see, e.g., Ref.~\cite{ich73}). The limit of the
vanishing magnetic field in Eq.~\eqref{eq:gamma3} is not trivial. An alternative but equivalent integral
form for the function \eqref{eq:gamma3} allowing easily the limit of the field-free case is derived in
Appendix~\ref{sec:app1}, see Eqs.~\eqref{eq:a15} and \eqref{eq:a16}. In Eq.~\eqref{eq:gamma2} the dynamical
polarization effects are included in a dielectric function $\varepsilon_{ei}(\mathbf{k},\omega )$. These effects guarantee
the convergence of the $\mathbf{k}$-integration in Eq.~\eqref{eq:gamma2} at large distances or at
small $\mathbf{k}$. But an upper cutoff $k_{\max}=1/r_{\min}$ (where $r_{\min}$ is the effective minimum
impact parameter) must be introduced in Eq.~\eqref{eq:gamma2} to avoid the logarithmic divergence at large
$\mathbf{k}$. This divergence corresponds to the incapability of the linearized kinetic equation to treat
close encounters between the plasma particles properly. Also it should be emphasized that for the e-i
collisions there are two specific frequencies $\gamma _{ie}=Z_{i} \gamma_{ei}$ and $\gamma _{ei}$ which
describe the relaxation of ionic and electronic temperatures to their equilibrium values, respectively
\cite{ich70,ich73}. Thus the total e-i collision frequency is given by $\widetilde{\gamma }_{ei}=%
\gamma _{ei}+ \gamma _{ie}=(Z_{i}+1)\gamma _{ei}$.

To estimate the range of variation of the collision frequency with increasing magnetic field consider now
some particular cases. At vanishing magnetic field from Eqs.~\eqref{eq:a15} and \eqref{eq:a16} we obtain
$G_{\alpha}(\mathbf{k},\omega )=(|k_{\parallel }|/k)\exp (-\omega ^{2}/2k^{2}v_{\alpha }^{2})$. In this limit
we denote $\gamma_{e\alpha}=\gamma_{0,e\alpha }$ with $\Lambda_{e\alpha}=\Lambda_{0,e\alpha }$ and from
Eq.~\eqref{eq:gamma2} we find
\begin{equation}
\ln \Lambda_{0,e\alpha }=\sqrt{\frac{2}{\pi }}\int_{0}^{k_{\max}}\frac{dk}{k} \int_{0}^{\infty }
\frac{e^{-u^{2}/2} u^{2}du}{\left\vert \varepsilon_{0,ei} \left(k,k\overline{v}_{e\alpha}u\right) \right\vert ^{2}}
\label{eq:h1}
\end{equation}
the Coulomb logarithm in the absence of a magnetic field. Here $\varepsilon_{0,ei}(k,\omega )$ is the usual longitudinal
dielectric function of the electron-ion unmagnetized TCP without collisions. Equation~\eqref{eq:h1}
has been derived by Ramazashvili \textit{et al}. \cite{ram62}. It involves the dynamical polarization effects through the
dielectric function $\varepsilon_{0,ei} (k,\omega )$ and requires only an upper cutoff $k_{\max} =1/r_{\min}$ in a Fourier
space. The Spitzer formula is recovered assuming $\varepsilon_{0,ei} (k,\omega )=1$ in Eq.~\eqref{eq:h1} and introducing
a lower cutoff $k_{\min} =1/\lambda_{\mathrm{D}}$. Then performing $u$-integration in Eq.~\eqref{eq:h1} we obtain the
usual Coulomb logarithm with $\Lambda_{0,e\alpha } =\lambda_{\mathrm{D}}/r_{e\alpha ,\min}$ generalized for
electron-ion plasmas (see, e.g., Ref.~\cite{bar06}). Here $\lambda _{\mathrm{D}}^{-2}=\lambda_{e}^{-2} +\lambda _{i}^{-2}$
while $r_{e\alpha ,\min}=\max[\lambda _{\mathrm{L}e\alpha};\lambda _{\mathrm{DB}}]$, where $\lambda_{\alpha}
=v_{\alpha}/\omega_{\alpha}$ and $\omega_{\alpha}=(4\pi n_{0\alpha}q_{\alpha}^{2}%
e^{2}/m_{\alpha})^{1/2}$ are the Debye screening length and the plasma frequency for plasma species $\alpha $,
respectively. $\lambda _{\mathrm{L}e\alpha}$ denotes the usual Landau length $\lambda _{\mathrm{L}ee}=
\lambda _{\mathrm{L}ei}/|Z_{i}|=e^{2}/3k_{B}T_{e}$ and $\lambda _{\mathrm{DB}}=\hslash /2\sqrt{m_{e}k_{B}T_{e}}$,
the electron de Broglie wavelength, taking care of the intrinsically quantum behavior of the high-temperature
plasma in the short range limit.

In the opposite case of an infinitely strong magnetic field ($\gamma_{e\alpha} =\gamma_{\infty ,e\alpha}$ with
$\Lambda_{e\alpha } =\Lambda_{\infty ,e\alpha}$) from Eqs.~\eqref{eq:a15} and \eqref{eq:a16}
one obtains $G_{\alpha }(\mathbf{k},\omega ) =\exp (-\omega ^{2}/2k_{\parallel }^{2}v_{\alpha }^{2})$. Inserting
this formula into Eq.~\eqref{eq:gamma2} it is straightforward to show that the collision frequency in a strong magnetic
field is the half of the frequency $\gamma_{0,e\alpha }$. Thus
\begin{equation}
\ln\Lambda_{\infty ,e\alpha }=\frac{1}{2}\ln\Lambda_{0,e\alpha } .
\label{eq:h7}
\end{equation}

The collision frequencies have been also investigated by some authors using the Fokker-Planck kinetic equation
with the Landau integral of the collisions \cite{mon74,sil98}. As stated above this approach neglects the
dynamical polarization effects. In Appendix~\ref{sec:app2} we show briefly that starting with Eqs.~\eqref{eq:gamma1}
and \eqref{eq:gamma2} where $\varepsilon_{ei} (\mathbf{k},\omega )$ is set $=1$, one arrives at the expressions
derived by Silin \cite{sil98}. To understand the importance of the dynamical polarization effects which
are neglected in Eqs.~\eqref{eq:b2} and \eqref{eq:b6} (and also in the formula of Spitzer) we note that the
kinetic equation in the form of Landau accounts for only close (almost) Coulomb collisions, thus completely neglecting
long-range wave-particle interactions. It has been shown previously that in the absence of magnetic field
corrections $\Delta_{e\alpha }$ to the standard Coulomb logarithm $\ln (\lambda_{\mathrm{D}}/r_{e\alpha ,\min})$
arising due to these interactions may be of the same order as the leading term \cite{ram62}. However, this effect is
crucial only for e-i interactions and wave-particle interactions are not expected to make any essential change
in the rate of e-e Coulomb collisions. For instance, in an anisothermic electron-ion plasma low-frequency
ion-acoustic waves may provide an effective mechanism for electron-ion interactions which leads to an enhancement
of the standard Coulomb logarithm.

\begin{figure*}[tbp]
\includegraphics[width=80mm]{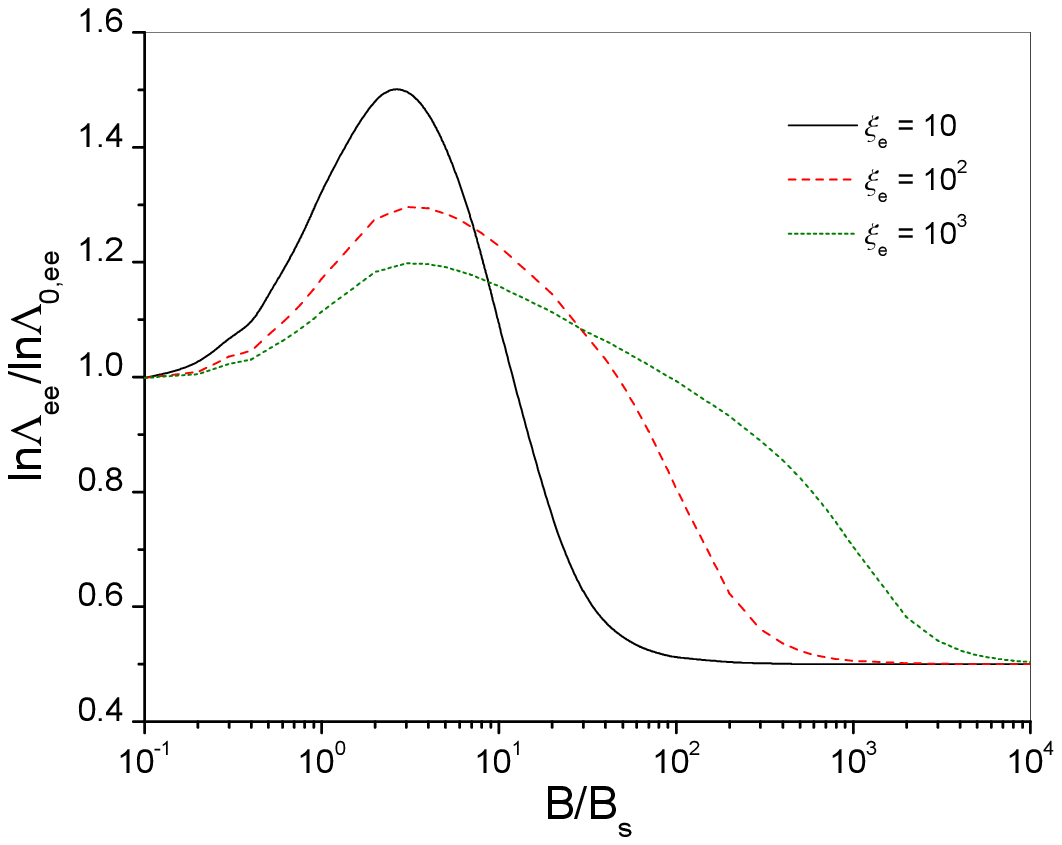}
\includegraphics[width=80mm]{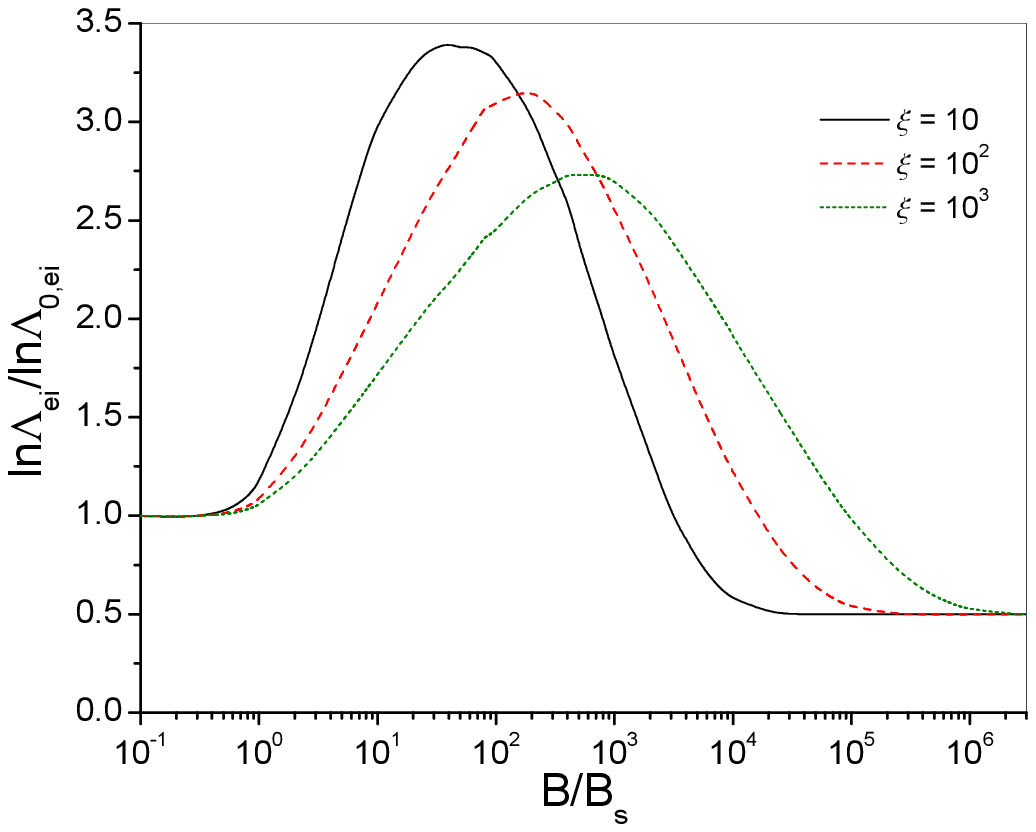}
\caption{(Color online) The Coulomb logarithms, Eqs.~\eqref{eq:b2} and \eqref{eq:b6}, normalized to the field-free values
for e-e (left panel) and electron-proton (right panel) collisions vs scaled magnetic field $B/B_{s}$ and for different
values of the cutoff parameter. See the text for explanations.}
\label{fig:2}
\end{figure*}

Similar long-range and low-frequency collective effects are responsible for a strong enhancement of the collision
rates in a magnetized plasma. In this case it has been shown \cite{ich70,ich73} that e-i collision rate contains, in
addition to the contribution from close Coulomb collisions \eqref{eq:b6}, a term which in our notations is given by
\begin{equation}
\ln\Lambda^{\ast}_{ei}=\frac{1}{4}\ln\left(\frac{m_{i}}{m_{e}}\right) e^{z} \left[\left(1+z\right)K_{0}\left(z\right)
-z K_{1}\left(z\right)\right] ,
\label{eq:an}
\end{equation}%
where $z=1/\zeta^{2}$, $\zeta =\lambda_{e}/a_{e}=\Omega_{e}/\omega_{e}$ is the scaled magnetic field, $K_{0}(z)$ and
$K_{1}(z)$ are the modified Bessel functions of the second kind. The Coulomb logarithm \eqref{eq:an} depends essentially
on the magnetic field. At large and small magnetic fields Eq.~\eqref{eq:an} behaves as $\ln\Lambda^{\ast}_{ei}\sim \ln
(\Omega_{e}/\omega_{e})$ and $\ln\Lambda^{\ast}_{ei}\sim \Omega_{e}/\omega_{e}$, respectively, and vanishes at
$\Omega_{e}\to 0$. Also we note the large factor $\ln (m_{i}/m_{e})$ in Eq.~\eqref{eq:an} which diverges at $m_{i}\to
\infty$ and appears due to strong electron-ion interactions via collective plasma waves and is typically $\gtrsim 10$.
However, Eq.~\eqref{eq:b6} which accounts for only close Coulomb collisions does not contain such term and remains finite
at $m_{i}\to \infty$.

\begin{table*}[tbp]
\caption{The Coulomb logarithm \eqref{eq:an} for electron-proton plasma normalized to $\ln \Lambda_{0,ei}$ for some
values of the scaled magnetic field $B/B_{s}$ and for different values of the cutoff parameter $\xi=\lambda_{\mathrm{D}}%
/r_{ei,\min}=10,10^{2},10^{3}$.}
\label{tab:1}
\begin{center}
\begin{tabular}{p{1.5cm}p{1.5cm}p{1.5cm}p{1.5cm}p{1.5cm}p{1.5cm}p{1.5cm}p{1.5cm}p{1.5cm}}
\hline\hline
$B/B_{s} $  & $0.1$ & $1.0$ & $10.0$ & $10^{2}$ & $10^{3}$ & $10^{4}$ & $10^{5}$ & $10^{6}$ \\
\hline
$\xi =10$     & 0.05 & 0.53 & 3.11 & 6.79 & 10.55 & 14.31 & 18.07 & 21.82 \\
$\xi =10^{2}$ & 0.03 & 0.27 & 1.55 & 3.40 &  5.28 &  7.16 &  9.03 & 10.91 \\
$\xi =10^{3}$ & 0.02 & 0.18 & 1.04 & 2.27 &  3.52 &  4.77 &  6.02 &  7.27 \\
\hline
\end{tabular}%
\end{center}
\end{table*}

The results of the numerical evaluation of Eqs.~\eqref{eq:b2} and \eqref{eq:b6} are shown in Fig.~\ref{fig:2}. This
figure shows the ratio $\ln\Lambda_{e\alpha }/\ln\Lambda_{0,e\alpha }$ as a function of the scaled magnetic field
$B/B_{s}$ for e-e (left panel) and electron-proton (right panel) collisions and for different values of the cutoff
parameters ($\xi_{e} =\lambda_{\mathrm{D}}/r_{ee,\min}$ (e-e) and $\xi =\lambda_{\mathrm{D}}/r_{ei,\min}$ (e-i)). The quantity
$B_{s}$ introduced above is $B_{s}=mcv_{e}/e\lambda_{\mathrm{D}}$. Also for an electron-proton plasma the normalized
Coulomb logarithm $\ln\Lambda^{\ast}_{ei}/\ln\Lambda_{0,ei}$, Eq.~\eqref{eq:an}, is given in Table~\ref{tab:1} for some
values of the scaled magnetic field. It is seen that the magnetic field may essentially increase the collision rates in
plasma compared to the field-free ones and this is more important for electron-ion collisions. As discussed above,
Eqs.~\eqref{eq:b2} and \eqref{eq:b6} and hence the results shown in Fig.~\ref{fig:2} account for only unscreened Coulomb
collisions neglecting dynamic polarization effects. In a vanishing magnetic field these effects are important for e-i
collisions and the situation with $B\neq 0$ requires further investigations, in particular for e-e collisions. In the
e-i case a major contribution is expected from low-frequency collective modes given by Eq.~\eqref{eq:an}. Table~\ref{tab:1}
shows that this contribution exceeds the Coulomb logarithm $\ln\Lambda_{ei}$ at $B/B_{s}\sim 10,10^{2},10^{3}$ depending
on the cutoff parameter. Also it should be emphasized that the validity of the regime \eqref{eq:h7} of a classically
strong magnetic field (the domains $B/B_{s}>10^{4}$ and $B/B_{s}>10^{6}$ in the left and right panels of Fig.~\ref{fig:2},
respectively) requires the condition $\hslash\Omega _{e}<k_{B}T_{e}$ (or $B<B_{c}=(mc/e\hslash )k_{B}T_{e}$). Thus the
results shown in Fig.~\ref{fig:2} are valid up to $B_{c}/B_{s}=k_{B}T_{e}/\hslash\omega_{e}$. Clearly the realization
of the regime \eqref{eq:h7} requires high temperatures and low densities and the enhancement of the collision rates
at $B\sim 1-10^{2}B_{s}$ may not be accessible under certain conditions. However the recent analysis shows \cite{kor08}
(see also the references therein) that in a quantizing magnetic field with $B>B_{c}$ the field-dependence of the
collisional rates becomes even stronger and the enhancement of $\gamma_{e\alpha }$ shown in Fig.~\ref{fig:2} may turn
even more significant although the classical expressions \eqref{eq:gamma1}, \eqref{eq:gamma2}, \eqref{eq:b2} and \eqref{eq:b6}
are invalid in this regime.

\section{Low-velocity stopping power}
\label{sec:s4}

In this section, with the collision-inclusive dielectric function derived in Sec.~\ref{sec:s2} we consider the stopping
power (SP) of a low-velocity ion moving in a magnetized plasma for an arbitrary angle with respect to the
magnetic field. The regime of low velocities is of particular importance for some physical situations, e.g., for electron
cooling processes \cite{ner07} and for magnetized target plasma fusion researches \cite{cer00}. Previously the SP in a
magnetized plasma at small ion velocity has been investigated by employing linear response (LR) theory \cite{ner00} and
Dufty-Berkovsky relation \cite{deu08}. The latter approach (see below) reduces the problem to a determination of the
diffusion coefficient of the magnetized plasma. In Ref.~\cite{ner00} it has been shown that in the presence of a magnetic
field and in the absence of collisions, the friction coefficient contains an anomalous term which diverges at $v\to 0$ like
$\ln (v_{e}/v)$ in addition to the usual constant one while the hydrodynamic approach of Ref.~\cite{deu08} does not contain
such term. We shall comment on this feature in this section.

The stopping power $S$ of an ion with charge $Ze$ and velocity $\mathbf{v}$ is defined as the energy loss of the ion in
a unit length due to interaction with a plasma. From Eq.~\eqref{eq:17} it is straightforward to calculate the electric
field $\mathbf{E}=-\nabla\varphi$ (or $\mathbf{E}(\mathbf{k},\omega )=-i\mathbf{k}\varphi (\mathbf{k},\omega)$ in terms
of Fourier transforms), and the stopping force acting on the ion. Then, the stopping power of the projectile pointlike
ion becomes (see, e.g., Refs.~\cite{ner00,ner03,ner98})
\begin{equation}
S=\frac{Z^{2}e^{2}}{2\pi^{2} v} \int d\mathbf{k} \frac{\mathbf{k}\cdot \mathbf{v}}{k^{2}}
\mathrm{Im} \frac{-1}{\varepsilon_{M}\left(\mathbf{k},\mathbf{k}\cdot \mathbf{v} ,\gamma\right)} .
\label{eq:st}
\end{equation}%
For the friction coefficient we have to consider $S$, given by Eq.~\eqref{eq:st} in a low-velocity limit, and thus
the dielectric function \eqref{eq:18} with \eqref{eq:23} and the functions $F_{1}(\mathbf{k},\omega )$ and $F_{2}(\mathbf{k},%
\omega )$ given by Eqs.~\eqref{eq:a11} and \eqref{eq:a12}, when $\omega =\mathbf{k}\cdot \mathbf{v}$. Now we have to write
the Taylor expansion of Eq.~\eqref{eq:18} for small $\omega =\mathbf{k}\cdot \mathbf{v}$. Using expressions \eqref{eq:18},
\eqref{eq:23}, \eqref{eq:21}, \eqref{eq:28}, \eqref{eq:a11}, and \eqref{eq:a12} for the collision-inclusive dielectric
function at $\omega\to 0$ we obtain
\begin{equation}
\mathrm{Im} \frac{-1}{\varepsilon _{M}\left( \mathbf{k},\omega ,\gamma \right) }\simeq
\frac{k\lambda _{e}^{2}}{\left(k^{2}\lambda _{e}^{2}+1\right) ^{2}}\frac{\omega }{v_{e}}
\int_{0}^{\infty}e^{-X\left( t\right) -\varsigma t}dt ,
\label{eq:38}
\end{equation}%
where $\varsigma =\gamma /kv_{e}=\nu /k\lambda_{e}$, $\nu =\gamma /\omega_{e}$. The function $X(t)$ is determined by
Eq.~\eqref{eq:a10}. It should be emphasized that Eq.~\eqref{eq:38} does not contain any logarithmic singularity at vanishing
$k_{\parallel}\to 0$ as for the case of collisionless magnetized plasma, see Ref.~\cite{ner00}. This singularity which leads
to an anomalous friction in a magnetized plasma has been removed here due to the collisions and the factor $e^{-\varsigma t}$
in Eq.~\eqref{eq:38} guarantees the convergence of the $t$ integration at $k_{\parallel}\to 0$. Thus from Eqs.~\eqref{eq:st}
and \eqref{eq:38} we obtain usual (linear with respect to $v$) friction law
\begin{equation}
S(\vartheta )\simeq \frac{2Z^{2}e^{2}}{\sqrt{2\pi }\lambda _{e}^{2}}\frac{v}{v_{e}} \mathcal{R}(\vartheta ) ,
\label{eq:39}
\end{equation}
where $\mathcal{R}(\vartheta )$ is the dimensionless friction coefficient,
\begin{equation}
\mathcal{R}(\vartheta )=\int_{0}^{\kappa }\frac{k^{3}dk}{\left( k^{2}+1\right) ^{2}}%
\left[\psi_{1}(k) \cos^{2}\vartheta + \frac{1}{2}\psi_{2}(k) \sin^{2}\vartheta\right] .
\label{eq:39-a}
\end{equation}%
Here $\kappa =k_{\max}\lambda_{e}$ and $\vartheta $ is the angle between $\mathbf{v}$ and $\mathbf{b}$. In Eq.~\eqref{eq:39-a}
we have introduced a cutoff parameter $k_{\max}=1/r_{\min}$ (where $r_{\min}$ is the effective minimum impact parameter) in
order to avoid the logarithmic divergence at large $k$. This divergence corresponds to the incapability of the linearized kinetic
theory to treat close encounters between the projectile ion and the plasma electrons properly. For $r_{\min}$ we thus use the
effective minimum impact parameter of classical binary Coulomb collisions which at low-velocities of the ion reads $r_{\min }%
=\vert Z\vert e^{2}/mv_{e}^{2}$. It is seen that the parameter $\kappa =4\pi n_{0e}\lambda^{3}_{e}/\vert Z\vert=1/\mathcal{Z}\gg 1$, where
$\mathcal{Z}$ is determined by Eq.~\eqref{eq:coup} at $v\ll v_{e}$. Also the other quantities in Eq.~\eqref{eq:39-a} are
\begin{equation}
\psi _{n}\left( k\right) =\frac{1}{2}\int_{0}^{\infty }\exp \left[ -\frac{%
2k^{2}}{\zeta ^{2}}\sin ^{2}\left( \zeta t\right) -2\nu t\right] \Phi
_{n}\left( kQ(t)\right) \frac{dt}{t\Upsilon \left( \zeta t\right) }
\label{eq:40}
\end{equation}%
with $n=1,2$, $\zeta =\lambda _{e}/a_{e}=\Omega_{e}/\omega_{e}$, $Q(t)=\sqrt{2}t\Upsilon \left( \zeta t\right) $,
$\Upsilon^{2}(t)=1-(\sin t/t)^{2}$, $\Phi_{1}(x) =x^{-2} \Phi (x)$, $\Phi_{2}(x) =2\mathrm{erf} (x)-x^{-2}\Phi (x)$.
The function $\Phi (x)$ is determined by Eq.~\eqref{eq:b5}.

In many experimental situations, the ions move in a plasma with random orientations of $\vartheta$ with respect
to the magnetic field direction $\mathbf{b}$. The friction coefficient appropriate to this situation may be obtained by
carrying out a spherical average over $\vartheta$ of $\mathcal{R}(\vartheta)$ in Eq.~\eqref{eq:39-a}. We find
\begin{equation}
\langle \mathcal{R}(\vartheta)\rangle =\frac{1}{3}\int_{0}^{\kappa }\frac{k^{3}dk}{\left( k^{2}+1\right) ^{2}}%
\left[\psi_{1} \left(k\right)+\psi_{2} \left(k\right)\right] .
\label{eq:39a}
\end{equation}

Let us analyze the general expression \eqref{eq:39-a} for some particular cases. For instance, at vanishing magnetic
field ($\zeta\to 0$) using the relation $Q(t)\simeq \sqrt{2/3}\zeta t^{2}$ at $\zeta\to 0$, one finds
\begin{equation}
\psi_{1}\left(k\right) =\frac{1}{2} \psi_{2}\left( k\right) =\frac{1}{3} A\left(
\frac{\nu }{\sqrt{2}k}\right) ,
\label{eq:41}
\end{equation}%
where $A (z) =e^{z^{2}} \mathrm{erfc} (z)$, $\mathrm{erfc} (z)$ is the complementary error function. In this case the
friction coefficient is isotropic and becomes
\begin{equation}
\mathcal{R}_{0}(\vartheta ) =\frac{1}{3}\int_{p_{0}}^{\infty }\frac{A \left( \nu k\right) dk}{k\left(
2k^{2}+1\right) ^{2}} .
\label{eq:43}
\end{equation}%
Here $1/p_{0}=\sqrt{2}\kappa $. In addition at vanishing damping, i.e. at $\nu \to 0$, $A (\nu k)\to 1$
and we recover the usual low-velocity stopping power in an unmagnetized collisionless plasma with a
friction coefficient (see, e.g., \cite{ner00,pet91})
\begin{equation}
\mathcal{R}_{0}(\vartheta )=\frac{1}{6} U(\kappa )\equiv \frac{1}{6}\left[\ln \left( 1+\kappa ^{2}\right)
-\frac{\kappa ^{2}}{\kappa ^{2}+1}\right] .
\label{eq:44}
\end{equation}

At strong magnetic field ($\zeta\to \infty$), the plasma becomes highly anisotropic and the friction coefficient
depends essentially on the angle $\vartheta$. For an evaluation of the functions $\psi_{1}(k)$ and $\psi_{2}(k)$ we
note that $Q(t)\to \sqrt{2}t$ and $\Upsilon (\zeta t)\to 1$ as $\zeta\to \infty$. Then substituting these relations
into Eq.~\eqref{eq:40} and after integration by parts one obtains
\begin{eqnarray}
&&\psi_{1}\left( k\right) =\frac{1}{2} A\left(a\right) +a^{2} B\left(a\right) -\frac{a}{\sqrt{\pi}} ,  \label{eq:45a} \\
&&\psi_{2}\left( k\right) =\left(1-a^{2} \right) B\left(a\right) -\frac{1}{2} A\left(a\right) +\frac{a}{\sqrt{\pi}}
\label{eq:45b}
\end{eqnarray}
with $a=\nu /\sqrt{2}k$, and
\begin{equation}
B\left(z\right) =\int_{z}^{\infty }\frac{dt}{t}A \left(t\right) =\frac{2z}{\sqrt{\pi }}\int_{0}^{\infty }
\ln \left( t+\sqrt{t^{2}+1}\right) e^{-z^{2}t^{2}}dt .
\label{eq:47}
\end{equation}%
Then the friction coefficient at infinitely strong magnetic field reads
\begin{eqnarray}
&&\mathcal{R}_{\infty}(\vartheta ) =\frac{1}{2}\int_{p_{0}}^{\infty }\frac{dk}{k\left( 2k^{2}+1\right) ^{2}}\bigg\{
\sin^{2}\vartheta \left[\left(1-\nu^{2}k^{2}\right)B\left(\nu k\right) -\frac{1}{2}\left(A\left(\nu k\right)
-\frac{2\nu k}{\sqrt{\pi}} \right)\right]   \label{eq:46}  \\
&&+\cos^{2}\vartheta \left[A \left(\nu k\right) -\frac{2}{\sqrt{\pi }}\nu k+2\nu ^{2}k^{2}B\left( \nu k\right) %
\right] \bigg\} .  \nonumber
\end{eqnarray}%
Similarly for the angular averaged friction coefficient we obtain
\begin{equation}
\langle\mathcal{R}_{\infty}(\vartheta )\rangle =\frac{1}{3}\int_{p_{0}}^{\infty }\frac{B\left(\nu k\right)dk}%
{k\left(2k^{2}+1\right) ^{2}} .
\label{eq:46x}
\end{equation}

The function $B(z)$ involved in Eq.~\eqref{eq:46} at small $z$ behaves as $B(z)\simeq \ln (1/z)-C/2$, where
$C=0.5772$ is the Euler's constant, and diverges logarithmically when $z\to 0$. Using asymptotic behavior of this
function it is straightforward to calculate from Eq.~\eqref{eq:46} the friction coefficient at vanishing $\gamma $.
In this limit and in the leading order we obtain
\begin{equation}
\mathcal{R}_{\infty}(\vartheta ) =\frac{1}{4}\left\{\sin^{2}\vartheta \left[\left(\ln \frac{\sqrt{2}\omega_{e}}{\gamma}%
-\frac{C+1}{2}\right) U\left(\kappa \right) +U_{1}\left(\kappa \right)\right]+ U\left(\kappa \right) \cos^{2}\vartheta \right\} ,
\label{eq:48}
\end{equation}%
where $U(\kappa )$ is given by Eq.~\eqref{eq:44}, and
\begin{equation}
U_{1}\left(\kappa \right)=U\left(\kappa \right) \ln\kappa -\frac{1}{4}\left[\ln^{2}\left(\kappa^{2}+1\right)-2\ln
\left(\kappa^{2}+1\right) \right] -\frac{1}{2}\mathrm{Li}_{2}\left(\frac{\kappa^{2}}{\kappa^{2}+1}\right) .
\label{eq:48x}
\end{equation}%
Here $\mathrm{Li}_{2}(z)$ is the dilogarithm function. Note that at large $\kappa\gg 1$, which is a requirement of a
weak ion-plasma coupling, the functions $U_{1}(\kappa )$ and $U(\kappa )$ can be approximated by $U_{1}(\kappa )\simeq
\ln^{2}\kappa -\pi^{2}/12$ and $U(\kappa )\simeq 2\ln\kappa -1$, respectively. It is seen that the first term in
Eq.~\eqref{eq:48} diverges logarithmically at vanishing $\gamma$. It can be shown that the general expression \eqref{eq:39-a}
with \eqref{eq:40} for the friction coefficient derived for arbitrary but finite magnetic field behaves similarly. This
is a consequence due to the magnetic field since the field-free result \eqref{eq:43} remains finite as $\gamma\to 0$
(see, e.g., Eq.~\eqref{eq:44}). The divergent term in Eq.~\eqref{eq:48} vanishes, however, when the ion moves along
the magnetic field ($\vartheta =0$). Then the friction coefficient is solely given by the last term of Eq.~\eqref{eq:48}.
In addition, the friction coefficient Eq.~\eqref{eq:48} for strong magnetic fields shows an enhancement for ions moving
transverse ($\vartheta =\pi /2$) to the magnetic field compared to the case of the longitudinal motion ($\vartheta =0$).
This effect is in agreement with particle-in-cell simulation results \cite{ner07}.

\begin{figure*}[tbp]
\includegraphics[width=80mm]{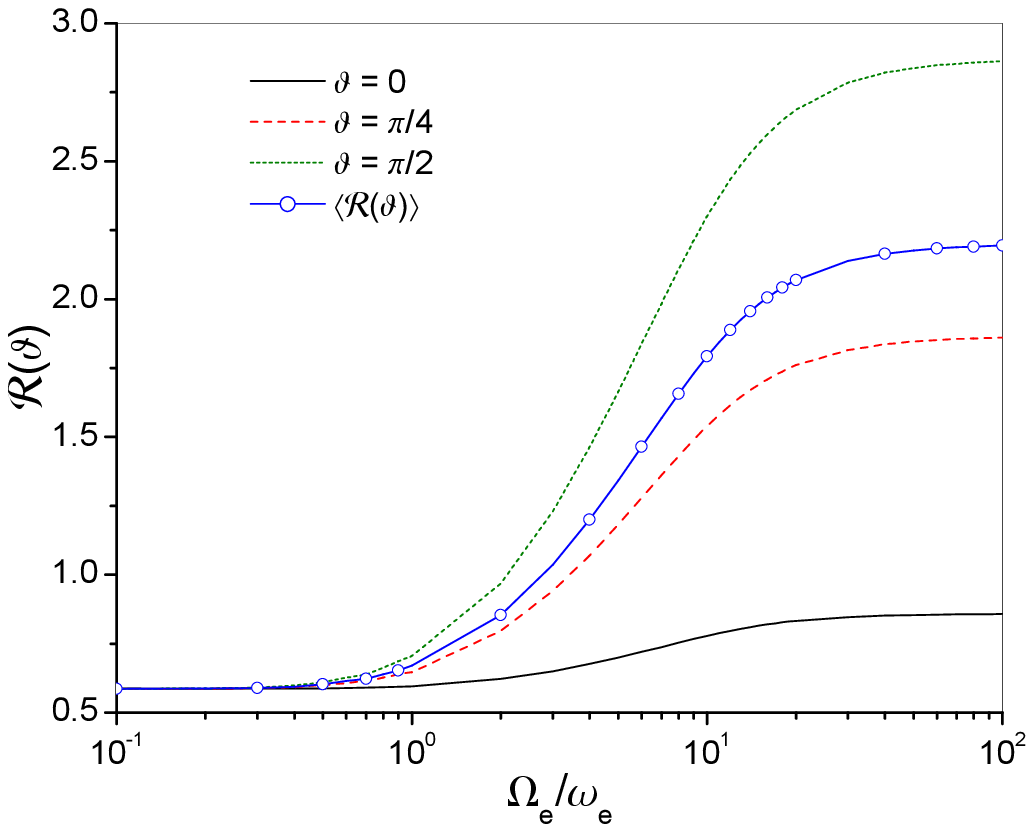}
\includegraphics[width=80mm]{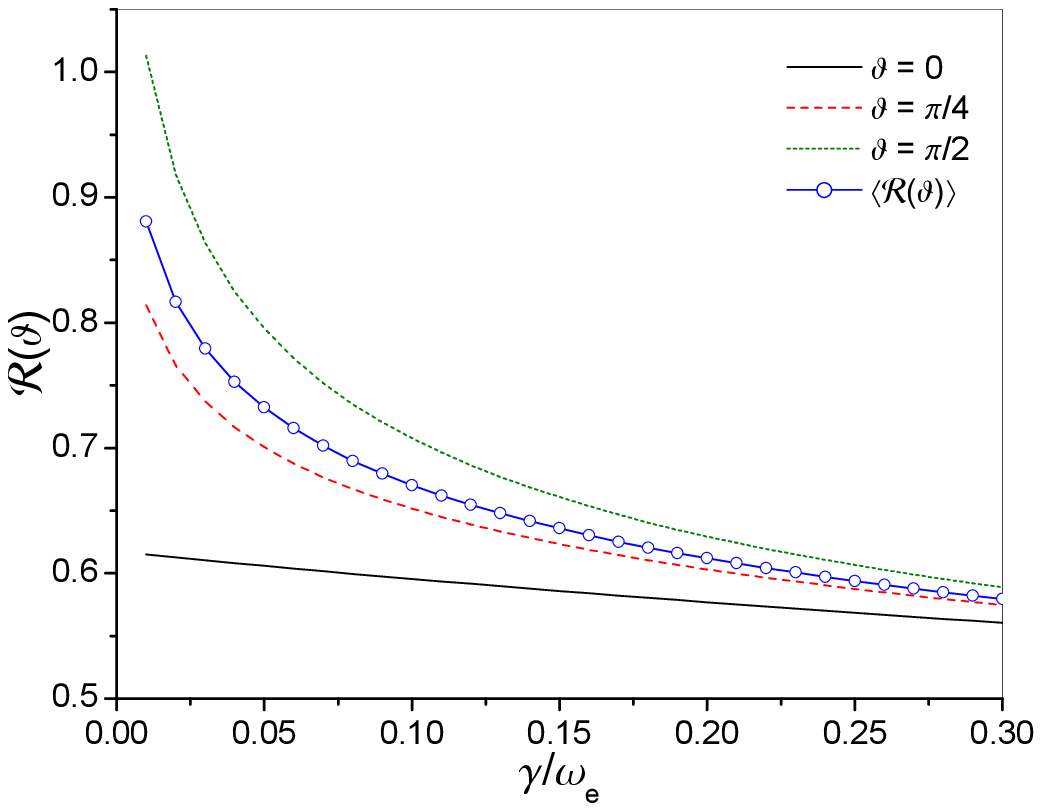}
\caption{(Color online) The friction coefficient $\mathcal{R}(\vartheta)$ vs the scaled magnetic field $\Omega_{e}/\omega_{e}$
(left panel) and damping parameter $\gamma /\omega_{e}$ (right panel) for $\vartheta =0$ (solid line), $\vartheta =\pi /4$
(dashed line) and $\vartheta =\pi /2$ (dotted line), $\kappa =10$, $\gamma /\omega _{e}=0.1$ (left panel), $\Omega_{e}=
\omega_{e}$ (right panel). The line with symbols corresponds to $\langle\mathcal{R}(\vartheta)\rangle$.}
\label{fig:3}
\end{figure*}

As stated in Introduction we shall now make contact with a different method. It has been shown by Dufty and Berkovsky
\cite{duf95} that the low-velocity SP of an ion in a plasma is related to the diffusion coefficient $D$ through
\begin{equation}
\left. \frac{S(v)}{v} \right\vert_{v\to 0} =\frac{k_{B}T_{e}}{D} .
\label{eq:db}
\end{equation}
As in Ref.~\cite{deu08} we consider $D$ to the self-diffusion coefficient in a magnetized classical one-component plasma.
From Eqs.~\eqref{eq:39}-\eqref{eq:40} we can relate the friction coefficient $\mathcal{R}(\vartheta )$ to the diffusion
coefficient $D$ through Eq.~\eqref{eq:db}. Cohen and Suttorp~\cite{coh84} have calculated parallel (to the magnetic field)
diffusion coefficient $D_{\parallel}$. These authors, like us but unlike Marchetti \emph{et al.}~\cite{mar87}, have used
a kinetic equation method. At vanishing damping ($\gamma\to 0$), it can be shown that $D_{\parallel}$ obtained from
Eqs.~\eqref{eq:39}-\eqref{eq:40} and for $\vartheta =0$ coincides with the result of Cohen and Suttorp~\cite{coh84}. In
particular, at $\gamma\to 0$, it is found from Eqs.~\eqref{eq:44} and \eqref{eq:48} that $\mathcal{R}_{0}(0)/\mathcal{R}_%
{\infty}(0) = D_{\infty ,\parallel}/D_{0,\parallel} =2/3$ in agreement with Ref.~\cite{coh84}. Here $D_{\infty ,\parallel}$
and $D_{0,\parallel}$ are the parallel self-diffusion coefficients at infinite and vanishing magnetic fields, respectively.
However at finite $\gamma$ and for $\vartheta =0$, comparing Eq.~\eqref{eq:43} with \eqref{eq:46} we conclude that the simple
relation cited above is not obeyed in general, due to damping.

As an example, we show in Fig.~\ref{fig:3} plots of the dimensionless friction coefficient $\mathcal{R}(\vartheta )$ given
by Eq.~\eqref{eq:39-a}, as a function of the scaled magnetic field $\Omega_{e}/\omega_{e}$ for model parameter $\gamma$
($\gamma /\omega_{e}=0.1$) (left panel). The right panel of Fig.~\ref{fig:3} shows $\mathcal{R}(\vartheta )$ as a function
of the scaled damping parameter $\gamma /\omega_{e}$ for $\Omega_{e}=\omega_{e}$ (i.e. for a given magnetic field). It is
seen that the low-velocity SP increases with an increase in the angle $\vartheta$ and also with the magnetic field. In the
latter case the SP asymptotically tends to the value given by expression~\eqref{eq:46}. In the opposite limit of a weak
magnetic field the friction coefficient tends to the value given by Eq.~\eqref{eq:43} which is independent of the angle
$\vartheta $. Also Fig.~\ref{fig:3} shows that the friction coefficient decreases with damping. It is, interestingly,
opposite to the behavior found for an unmagnetized DEG, see. e.g., Refs.~\cite{ner08,ner02,ash80,abr94}.
A decrease of $\mathcal{R}(\vartheta )$ with $\gamma$ in the present case of a classical plasma is not attributable to the
applied magnetic field because the field-free friction coefficient given by \eqref{eq:43} shows a similar behavior (not
shown in Fig.~\ref{fig:3}). In a degenerate plasma an enhancement of the low-velocity SP with $\gamma$ is a quantum effect
which is absent in our present study. For a DEG the domain of plasmon excitations is shifted towards smaller ion velocities
\cite{ner08,ner02}; this increases the SP in this velocity regime. But in the present case the domain of collective excitations is
shifted towards higher velocities \cite{pet91} and the friction coefficient decreases with $\gamma$.

The resulting friction coefficient \eqref{eq:48} may be compared with Eq.~(44) of Ref.~\cite{ner00} where the friction
coefficient in the collisionless plasma contains an anomalous term $\ln (v_{e}/v)$ vanishing at $\vartheta =0$. The
physical origin of such an anomalous friction coefficient may be traced to the spiral motion of the electrons
along the magnetic field lines. These electrons naturally tend to couple strongly with long-wavelength fluctuations
(i.e., small $k_{\parallel}$) along the magnetic field. In addition, when such fluctuations are characterized by slow
variation in time (i.e., small $\omega =\mathbf{k}\cdot \mathbf{v}$), the contact time or the rate of energy exchange
between the electrons and the fluctuations will be further enhanced. In a plasma, such low-frequency fluctuations
are provided by a slow projectile ion. The above coupling can therefore be an efficient mechanism of energy exchange
between the electrons and the projectile ion. At vanishing damping and in the limit of $v\to 0$, the frequency $\omega
=\mathbf{k}\cdot \mathbf{v} \to 0$ tends to zero as well. The contact time $\sim\omega^{-1}$ thus becomes infinite and
the friction coefficient diverges. The collisions of the plasma particles play a stabilizing role since the fluctuations
provided by the slow ion are damped. Thus at small velocities $v\to 0$ the contact time is finite and is determined by
$\sim\gamma^{-1}$. As a result Eq.~\eqref{eq:48} does not contain a term like $\ln (v_{e}/v)$ but behaves as $\ln (1/\gamma)$
at vanishing damping.

\section{Kinetic versus hydrodynamic approach}
\label{sec:s5}

Using the theoretical results obtained in Sec.~\ref{sec:s4}, we present here some comparative analysis, looking for some
contacts between our linear-response (kinetic) formulation and the previous hydrodynamic mode-coupling treatments based on the
self-diffusion coefficients. As mentioned above such connection is established via Dufty-Berkovsky relation \eqref{eq:db}.
The plasma is modeled as a collisional dielectric medium whose linear response function, within RTA, is given by Eqs.~\eqref{eq:18},
\eqref{eq:23}-\eqref{eq:25} with $\gamma$ as a model damping parameter. In order to document the LIVSD physics highlighted
by the relation \eqref{eq:db}, we first briefly pay attention to the unmagnetized $B=0$ limit. We consider it through
the small $\epsilon=1/(4\pi n_{0e}\lambda_{e}^{3})\ll 1$ plasma parameter approximation for the self-diffusion coefficient
given by Sj\"{o}gren \emph{et al}. \cite{sjo81}. Employing Eqs.~\eqref{eq:39}, \eqref{eq:44} and \eqref{eq:db} an inspection
shows that at vanishing damping ($\gamma =0$) the self-diffusion coefficient obtained from these formulas coincides with
the result of Ref.~\cite{sjo81} if the ionic charge number square $Z^{2}$ in Eq.~\eqref{eq:39} is replaced by the quantity
$Z^{2}\to P(Z)$, where
\begin{equation}
P(Z)=\left(Z+\frac{1}{\sqrt{2}}\right) \frac{32Z^{2}+75\sqrt{2}Z+50}{104Z^{2}+111\sqrt{2}Z+59} .
\label{eq:dif0}
\end{equation}
For a proton projectile with $Z=1$ this factor is $P(1)=1.003$ and the agreement between both approaches is almost perfect.
The factor \eqref{eq:dif0} which is nonlinear with respect to $Z$ accounts for the nonlinear coupling between an incoming ion
and the surrounding plasma \cite{sjo81}. However for highly charged ions with $Z\gg 1$ this factor increases linearly with
$Z$, $P(Z)=(4/13)Z$, while more rigorous treatment shows that at strong ion-plasma coupling the energy loss of an ion scales
with its charge approximately like $Z^{1.5}$ \cite{zwic99}.

Consider next the case of a magnetized and collisional plasma. For simplicity we consider electron-proton plasma and a
proton as a projectile particle. Exploring first the moderately magnetized domain, $\Omega_{e}\geqslant \omega_{e}$, one
can explicit the field-free parallel and $B$-dependent transverse diffusions \cite{mar87},
\begin{equation}
D_{\parallel }^{(0)}=\frac{3\sqrt{\pi }v_{p}^{2}}{\gamma _{c}},  \quad
D_{\perp }^{(0)}=\frac{r_{L}^{2}\gamma _{c}}{3\sqrt{\pi }} ,
\label{eq:dif1}
\end{equation}
where $v_{p}^{2}=k_{B}T_{e}/m_{p}$, $m_{p}$ is the proton mass, and $\gamma_{c}=\omega_{e}\epsilon\ln (1/\epsilon )$ is the
collision frequency in terms of the plasma parameter $\epsilon$, $r_{L}=v_{p}/\Omega_{e}$ is the Larmor radius. Note that
the collision frequency $\gamma_{c}$ is related to the e-e collisional relaxation rate $\gamma_{ee}$ as $\gamma_{ee} =%
\sqrt{2/9\pi}\gamma_{c}$ (see Sec.~\ref{sec:s3}). The transverse diffusion coefficient given by Eq.~\eqref{eq:dif1} corresponds
to a classical region, where $D_{\perp }\sim B^{-2}$, and is valid for $\gamma_{c}<\Omega_{e} <0.4\omega_{e} Y(\epsilon )$
with $Y(\epsilon )=[\epsilon^{2}\ln (1/\epsilon)]^{-1/2}$ as explained in Ref.~\cite{mar87}.

\begin{figure*}[tbp]
\includegraphics[width=80mm]{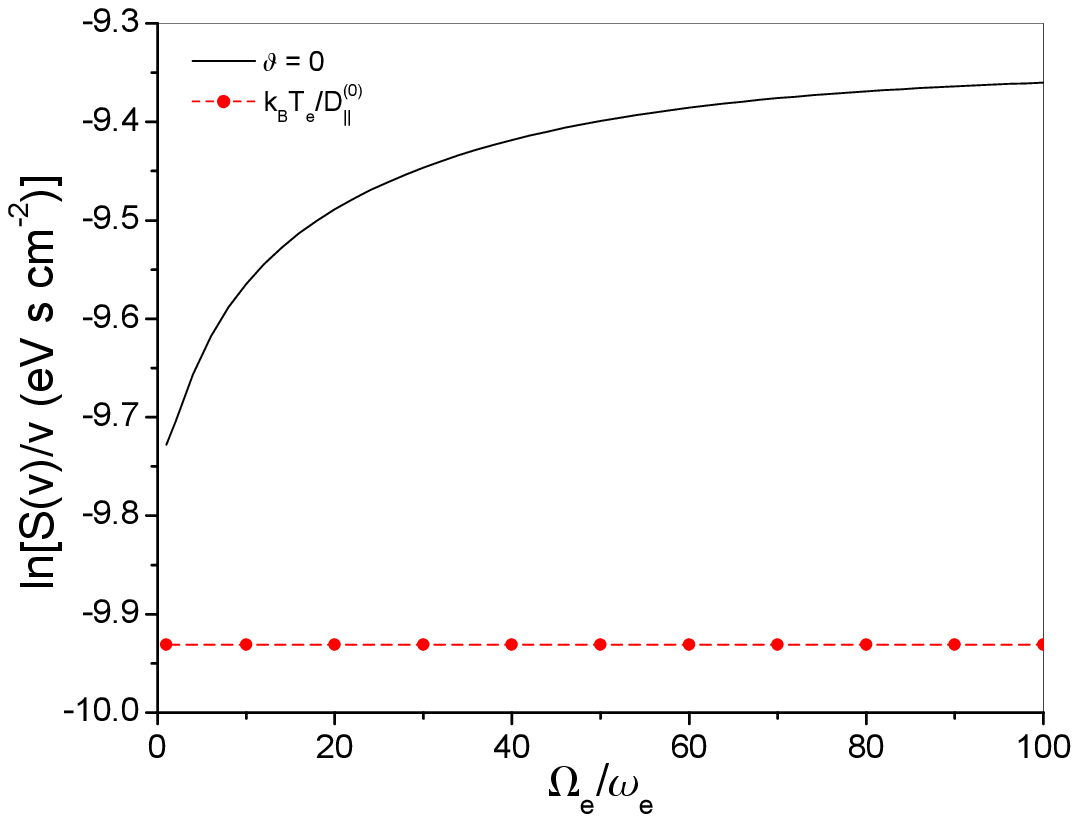}
\includegraphics[width=80mm]{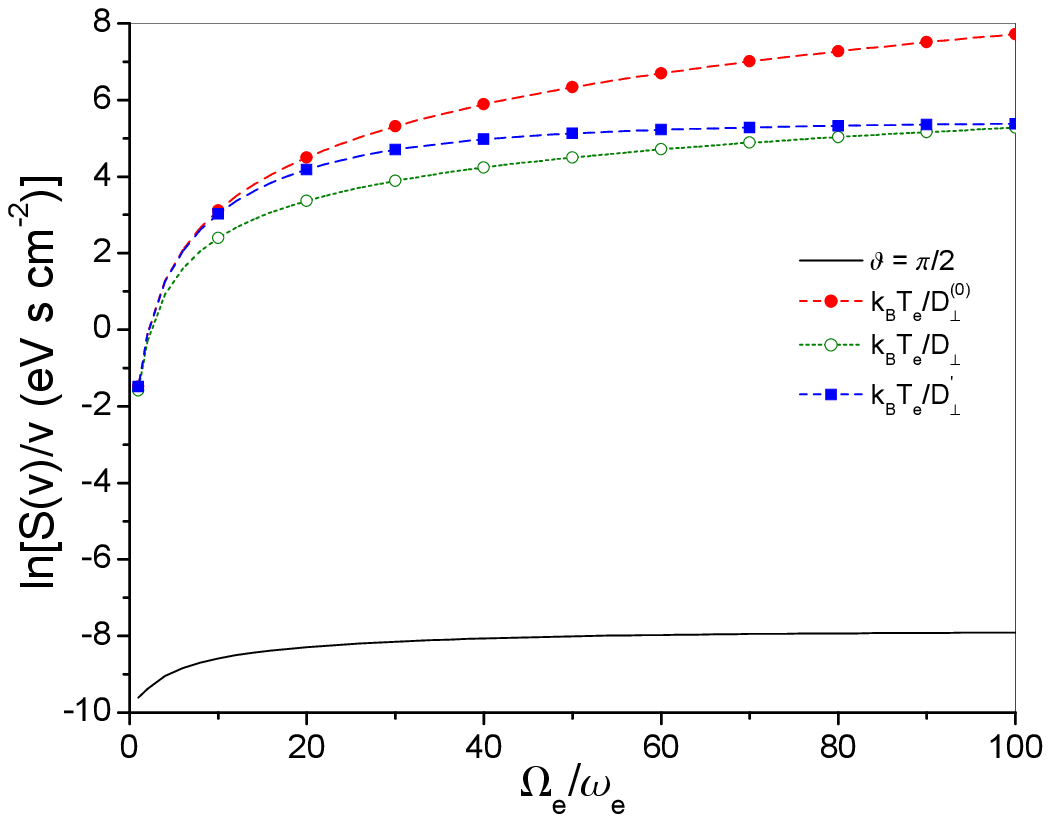}
\caption{(Color online) Proton LIVSD in a plasma with $n_{e}=1.064\times 10^{16}$ cm$^{-3}$, $T_{e}=1$ eV ($\epsilon =0.02$) in
terms of $\Omega_{e}/\omega_{e}$. The lines with symbols in left and right panels represent parallel and transverse LIVSD, respectively.
The solid lines were obtained from Eqs.~\eqref{eq:39}-\eqref{eq:40} with $\vartheta =0$ (left) and $\vartheta =\pi /2$ (right).}
\label{fig:4}
\end{figure*}

With higher magnetic field values, $\Omega_{e}/\omega_{e} >4Y(\epsilon )$, one reaches the transverse hydro-Bohm regime
with $D_{\perp }\sim B^{-1}$ featuring \cite{mar87}
\begin{equation}
D_{\perp} =D_{\perp}^{(0)}+\frac{v_{p}^{2}}{2\Omega _{e}}\epsilon^{2}\left[\ln
(1/\epsilon ) \right]^{3/2} .
\label{eq:dif2}
\end{equation}
To the intermediate plateau regime with $D_{\perp}\sim B^{0}$ between transverse diffusion coefficients given by
Eqs.~\eqref{eq:dif1} and \eqref{eq:dif2}, corresponds the diffusion coefficient \cite{mar87} valid at $0.4Y(\epsilon )
<\Omega_{e}/\omega_{e} <Y(\epsilon )$,
\begin{equation}
D^{\prime}_{\perp} =D_{\perp}^{(0)}\left[1+\frac{0.6\epsilon\gamma_{c}}{\omega_{e}} \zeta^{2}\right]
\label{eq:dif4}
\end{equation}
with $\zeta =\Omega_{e}/\omega_{e}$. When electron diffusion is considered, the electron thermal velocity $v_{e}$ should
be used in Eqs.~\eqref{eq:dif1}-\eqref{eq:dif4} instead of $v_{p}$. It is also important to stress that the quantitative
predictions \eqref{eq:dif1}-\eqref{eq:dif4} of the mode coupling theory developed in Ref.~\cite{mar87} are strongly
dependent on the values of the hydrodynamic cutoffs which, in contrast to the kinetic theory, are introduced linearly.
A reliable estimate of the magnetic field and plasma parameter dependence of the cutoffs have been obtained, but their
exact values are not known. Consequently, the numerical coefficients in Eqs.~\eqref{eq:dif1}-\eqref{eq:dif4} are not
precisely known. The exact coefficients can in principle be obtained by using kinetic theory.

The friction coefficients $S(v)/v$ (at $v\to 0$) calculated with the help of Eqs.~\eqref{eq:db} and
\eqref{eq:dif1}-\eqref{eq:dif4} are shown in Fig.~\ref{fig:4} as the lines with symbols. In this figure the solid lines
without symbols demonstrate the friction coefficients calculated from Eqs.~\eqref{eq:39}-\eqref{eq:40} with $\vartheta =0$
(left panel) and $\vartheta =\pi /2$ (right panel) assuming, for consistancy, the same collision frequency $\gamma =\gamma_{c}$
as in Eqs.~\eqref{eq:dif1}-\eqref{eq:dif4}. As far as we can see, there are no fundamental contradictions between kinetic
[Eqs.~\eqref{eq:39}-\eqref{eq:40}] and hydrodynamic [Eq.~\eqref{eq:db} and the first relation of Eq.~\eqref{eq:dif1}]
approaches for the parallel case, see Fig.~\ref{fig:4}, left panel. However, there are differences between the two approaches.
It is well known that a kinetic equation approach contains more information about a physical system than a hydrodynamical
approach. Indeed assuming a vanishing damping ($\gamma =0$) and
magnetic field ($B=0$) from Eqs.~\eqref{eq:39} and \eqref{eq:44} at $\epsilon\ll 1$ the ratio of the low-velocity SPs of
both approaches is $S_{\parallel ,\mathrm{kin}}/S_{\parallel ,\mathrm{hyd}}\simeq \sqrt{2}$. As discussed above the
numerical coefficients in Eqs.~\eqref{eq:dif1}-\eqref{eq:dif4} are not precisely known. Thus including the numerical factor
$\sqrt{2}$ into denominator of the parallel diffusion coefficient in Eq.~\eqref{eq:dif1} the agreement between both
approaches becomes complete. Note that this is equivalent to the redefinition $D_{\parallel}\to D^{\ast}_{\parallel}
=v^{2}_{p}/\gamma_{ee}$, where $\gamma_{ee}=\sqrt{2/9\pi}\gamma_{c}$.

An apparently large discrepancy is documented for the transverse situation (Fig.~\ref{fig:4}, right panel) where typically
$S_{\bot ,\mathrm{kin}}/S_{\bot ,\mathrm{hyd}}\sim [\epsilon\ln (1/\epsilon)]^{\alpha} \ll 1$ with $\epsilon\ll 1$ and
$\alpha$ varies between $2\leqslant\alpha\leqslant 4$ depending on the strength of $B$. The kinetic regime seems to be
restricted to $0.08=\gamma_{c}/\omega_{e}< \Omega_{e}/\omega_{e}<0.4Y(\epsilon )=10.1$ as explained in Ref.~\cite{mar87}.
The discrepancy in the orthogonal case might be due to a different treatment of cutoffs in kinetic and hydrodynamic theories, i.e
logarithmic vs linear. Actually, according to Ref.~\cite{mar87}, the different hydro modes are normalized to distinct
cutoffs. Upper hybrid ones are normalized to $1/a_{e}$, $a_{e}$ being electron Larmor radius while low frequency
modes are normalized to inverse mean free path $1/\ell$ with $\ell =v_{e}/\gamma_{c}$. On the other hand in the extreme
limit $\Omega_{e}\gg \omega_{e}$, the only reasonable transverse cutoff should be $1/a_{e}$ which, for instance, in the
kinetic treatment is included as $\ln (1/a_{e})$. The basic physics involved in this orthogonal geometry pertains to kinetic
theory when we rely on a collisional time while in the $B\to \infty$ limit leading to hydrodynamics, we incorporate the
Larmor rotation of the charged particles, as well.

\section{Discussion and outlook}
\label{sec:disc}

In this paper, we have presented a detailed investigation of the stopping power (SP) of low-velocity ions in a magnetized
and collisional classical plasma. In the course of this study we have derived, among other things, some analytical results
for the collision-inclusive linear response function for which the effect of collisions is taken into account in a
number-conserving manner relaxation-time approximation (RTA) based on the Boltzmann-Poisson equations--the BGK approach.
These analytical results at small projectile ion velocities go beyond those obtained in Refs.~\cite{ner98,ner00,ner03,see98,wal00,ste01}.
After a quantitative introduction to the linear response function in Sec.~\ref{sec:s2}, we have briefly studied the
effect of magnetic field on relaxation rates due to collisions between different plasma species in Sec.~\ref{sec:s3}
in which we have estimated contributions of close Coulomb collisions and long-range particle-wave interactions in the
presence of a magnetic field on $\gamma_{\alpha \beta}$. It has been shown that a magnetic field leads to an essential
increase in collision rates.

Theoretical calculations of SP based on the linear response theory within RTA are discussed in Sec.~\ref{sec:s4}. A
number of limiting and asymptotic regimes of low-velocities and vanishing $\gamma$ have been studied. The theoretical
expressions for SP derived in this section lead to a detailed presentation, in Secs.~\ref{sec:s4} and \ref{sec:s5},
of a collection of data through figures on SP of an ion. The results we have presented demonstrate that with
regard to SP the difference between RTA and the usual linear response theory without damping is substantial. In particular, we
have shown that the anomalous friction coefficient which behaves as $S_{\mathrm{an}}(v)/v \sim\ln (v_{e}/v)$ at $v\ll v_{e}$
obtained for a collisionless plasma \cite{ner00} is now absent. Such a term arises due to an enhancement of energy exchange
between plasma particles at $v\to 0$ when the contact time $\sim \omega^{-1}$ with $\omega\sim \mathbf{k}\cdot \mathbf{v}$ becomes
very large. Thus collisions in a plasma play a stabilizing role for the energy exchange process and in the case of a collisional
plasma the contact time is given by $\sim \gamma^{-1}$. For low-velocity SP this yields $S(v)/v\sim \ln (1/\gamma)$. Finally,
in Sec.~\ref{sec:s4} we have related the friction coefficient $\mathcal{R}(\vartheta )$ at $\vartheta =0$ to the parallel
diffusion coefficient $D_{\parallel}$ via Eq.~\eqref{eq:db} and at vanishing damping ($\gamma\to 0$), it has been shown that
$D_{\parallel}$ obtained from Eqs.~\eqref{eq:39}-\eqref{eq:40} coincides with the result of Ref.~\cite{coh84}. However, at
finite $\gamma$ the equivalence of both approaches is violated due to damping.

In Sec.~\ref{sec:s5} we have also compared the results obtained within our kinetic (Boltzmann-Poisson equations) approach with a
simple LIVSD expression~\eqref{eq:db}, using transverse and parallel diffusion coefficients derived within a hydrodynamic
mode-coupling theory \cite{mar87}. In the parallel case our results agree perfectly with this approach. There are discrepancies
in the transverse situation and these have been discussed in Sec.~\ref{sec:s5}. In the transverse case one may require an improved
mode-coupling calculation.

Going beyond the BGK approach which is based on the Boltzmann-Poisson equations we can envisage a number of avenues. These include
(i) extending the number-conserving RTA to number-, momentum- and energy-conserving RTA. We note that this has been done previously
in Ref.~\cite{mor00} for a field-free case; (ii) studying SP in a degenerate electron gas (DEG) in the presence of a non-quantizing
magnetic field, in a number-conserving RTA along the approach of Mermin \cite{mer70} and Das \cite{das75} neither of whom considered
a magnetic field; (iii) ion interaction with fluctuating plasma microfield and stochastic energy loss which may lead to a DB-like
relation from first principles, and (iv) improved mode-coupling hydrodynamics which will result in an energy loss calculation via
DB relation and also exploring rigorous kinetic calculations of transport coefficients. Lastly, it may be mentioned that a
quasiclassical model was studied by Das \cite{das76} which uses the Boltzmann-Poisson equations but also includes Landau quantization
in a DEG.

\begin{acknowledgments}
It is our pleasure to thank Prof. C. Toepffer and Dr. G. Zwicknagel for many stimulating conversations.
The work of H.B.N. has been partially supported by the Armenian Ministry of Higher Education and Science
(Project No. 87).
\end{acknowledgments}

\appendix

\section{Integral representation of the dielectric function}
\label{sec:app1}

In Sec.~\ref{sec:s2} we have derived the Bessel-function representation of the dielectric function. For some
applications an integral representation of the dielectric function is desirable. For deriving the integral form
of the dielectric function we rewrite the denominators of Eqs.~\eqref{eq:19} and \eqref{eq:20} using an integral
\begin{equation}
\frac{1}{\Omega -i\gamma }=i\int_{-\infty }^{0}e^{i\left( \Omega -i\gamma
\right) t}dt
\label{eq:a1}
\end{equation}%
and summation formula $\sum_{n=-\infty }^{\infty}J_{n}^{2}(z)e^{int} =J_{0}(2z\sin \frac{t}{2})$ \cite{gra80}.
Then the dielectric function may be alternatively represented in the form
\begin{eqnarray}
\varepsilon \left( \mathbf{k},\omega ,\gamma \right)  &=&1-\frac{\omega
_{e}^{2}}{k^{2}}\frac{2\pi }{n_{0e}}\int_{-\infty }^{\infty }dv_{\parallel
}\int_{0}^{\infty }v_{\perp }dv_{\perp }\int_{0}^{\infty }e^{-\gamma
t}e^{it\left( \omega -k_{\parallel }v_{\parallel }\right) }dt  \label{eq:a3} \\
&&\times \left[ ik_{\parallel }\frac{\partial f_{0}}{\partial v_{\parallel }}%
J_{0}\left( w\left( t\right) \right) +k_{\perp }\frac{\partial f_{0}}{%
\partial v_{\perp }}\cos \frac{\Omega _{e}t}{2}J_{1}\left( w\left( t\right)
\right) \right] ,  \nonumber \\
Q\left( \mathbf{k},\omega ,\gamma \right)  &=&\frac{2\pi }{n_{0e}}%
\int_{-\infty }^{\infty }dv_{\parallel }\int_{0}^{\infty }f_{0}\left(
v_{\parallel },v_{\perp }\right) v_{\perp }dv_{\perp }\int_{0}^{\infty
}e^{-\gamma t}e^{it\left( \omega -k_{\parallel }v_{\parallel }\right) }dt
\label{eq:a4} \\
&&\times \left[ ik_{\parallel }v_{\parallel }J_{0}\left( w\left( t\right)
\right) +k_{\perp }v_{\perp }\cos \frac{\Omega _{e}t}{2}J_{1}\left( w\left(
t\right) \right) \right] ,  \nonumber
\end{eqnarray}%
where the argument of the Bessel functions is given by
\begin{equation}
w\left( t\right) =\frac{2k_{\perp }v_{\perp }}{\Omega _{e}}\sin \frac{\Omega
_{e}t}{2} .
\label{eq:a5}
\end{equation}%
In the above expressions by performing integration by parts, one finally obtains
\begin{eqnarray}
\varepsilon \left( \mathbf{k},\omega ,\gamma \right)  &=&1+\frac{\omega
_{e}^{2}}{k^{2}}\frac{2\pi }{n_{0e}}\int_{-\infty }^{\infty }dv_{\parallel
}\int_{0}^{\infty }f_{0}\left( v_{\parallel },v_{\perp }\right) v_{\perp
}dv_{\perp }  \label{eq:a6} \\
&&\times \int_{0}^{\infty }e^{-\gamma t}e^{it\left( \omega -k_{\parallel
}v_{\parallel }\right) }J_{0}\left( w\left( t\right) \right) \left[
k_{\parallel }^{2}+k_{\perp }^{2}\frac{\sin \left( \Omega _{e}t\right) }{%
\Omega _{e}t}\right] tdt,  \nonumber \\
Q\left( \mathbf{k},\omega ,\gamma \right)  &=&1+\frac{2\pi i}{n_{0e}}\left(
\omega +i\gamma \right) \int_{-\infty }^{\infty }dv_{\parallel
}\int_{0}^{\infty }f_{0}\left( v_{\parallel },v_{\perp }\right) v_{\perp
}dv_{\perp }  \label{eq:a7} \\
&&\times \int_{0}^{\infty }e^{-\gamma t}e^{it\left( \omega -k_{\parallel
}v_{\parallel }\right) }J_{0}\left( w\left( t\right) \right) dt.  \nonumber
\end{eqnarray}%
In particular, for a Maxwellian isotropic distribution function \eqref{eq:22}, Eqs.~\eqref{eq:a6}
and \eqref{eq:a7} become
\begin{eqnarray}
&&\varepsilon \left( \mathbf{k},\omega ,\gamma \right) =1+\frac{1}{%
k^{2}\lambda _{e}^{2}}\left[ 1+\left( is-\varsigma \right) \int_{0}^{\infty
}e^{ist-X\left( t\right) -\varsigma t}dt\right] ,  \label{eq:a8} \\
&&Q\left( \mathbf{k},\omega ,\gamma \right) =1+\left( is-\varsigma \right)
\int_{0}^{\infty }e^{ist-X\left( t\right) -\varsigma t}dt,  \label{eq:a9}
\end{eqnarray}%
where $s=\omega /kv_{e}$, $\varsigma =\gamma /kv_{e}$, and
\begin{equation}
X\left(t\right) =\frac{t^{2}}{2}\frac{k_{\parallel }^{2}}{k^{2}}+k_{\perp
}^{2}a_{e}^{2}\left[ 1-\cos \left( \frac{t}{ka_{e}}\right) \right] .
\label{eq:a10}
\end{equation}%
Here $a_{e}=v_{e}/\Omega _{e}$ is the cyclotron radius of the electrons. The real and imaginary parts of
the plasma dispersion function defined by Eq.~\eqref{eq:23} can be found from Eq.~\eqref{eq:a8} and are
given by
\begin{eqnarray}
&&F_{1}\left( \mathbf{k},\omega \right) =1-\int_{0}^{\infty }\left[ s\sin
\left( st\right) +\varsigma \cos \left( st\right) \right] e^{-X\left( t\right)
-\varsigma t}dt,  \label{eq:a11} \\
&&F_{2}\left( \mathbf{k},\omega \right) =\int_{0}^{\infty }\left[ s\cos
\left( st\right) -\varsigma \sin \left( st\right) \right] e^{-X\left( t\right)
-\varsigma t}dt .  \label{eq:a12}
\end{eqnarray}

In this section we also derive an alternative integral representation for the collision frequency
\eqref{eq:gamma1} in a magnetized plasma. Using an integral
\begin{equation}
e^{-u^{2}}=\frac{1}{\sqrt{\pi }}\int_{-\infty }^{\infty }e^{2iut-t^{2}}dt
\label{eq:a13}
\end{equation}%
and the summation formula (see, e.g., Ref.~\cite{gra80})
\begin{equation}
\sum_{n=-\infty }^{\infty }\Lambda _{n}\left( z\right) e^{-in\theta }=\exp
\left[ -z\left( 1-\cos \theta \right) \right]
\label{eq:a14}
\end{equation}
from Eq.~\eqref{eq:gamma3} we obtain
\begin{equation}
G_{\alpha }\left( \mathbf{k},\omega \right) =\sqrt{\frac{2}{\pi }}\left\vert
k_{\parallel }\right\vert v_{\alpha }\int_{0}^{\infty }e^{-U_{\alpha }\left(
\mathbf{k},t\right) }\cos \left( \omega t\right) dt ,
\label{eq:a15}
\end{equation}%
where
\begin{equation}
U_{\alpha }\left( \mathbf{k},t\right) =k_{\parallel }^{2}v_{\alpha }^{2}%
\frac{t^{2}}{2}+k_{\perp }^{2}a_{\alpha }^{2}\left[ 1-\cos \left( \Omega
_{\alpha }t\right) \right] .
\label{eq:a16}
\end{equation}%
Let us recall that here $a_{\alpha }=v_{\alpha }/\Omega _{\alpha }$, $v_{\alpha}$, and $\Omega_{\alpha}$ are
the cyclotron radius, the thermal velocity and the cyclotron frequency of the plasma species $\alpha $,
respectively.

\section{Collision frequencies obtained from the Landau kinetic equation}
\label{sec:app2}

In this appendix we show that Eq.~\eqref{eq:gamma2} for the Coulomb logarithm is equivalent to the
formula obtained by Silin in Ref.~\cite{sil98} if one neglects the dynamical polarization of the
plasma i.e. assuming $\varepsilon _{ei}(\mathbf{k},\omega)=1$ in Eq.~\eqref{eq:gamma2}. Note that the
latter approach corresponds to the kinetic equation with the collisional integral in the form of Landau
generalized for the case of a magnetized plasma.

Using the integral representation~\eqref{eq:a15} of the function $G_{\alpha}(\mathbf{k},\omega)$
we obtain
\begin{equation}
\int_{-\infty }^{\infty }G_{e}\left( \mathbf{k},\omega \right)
G_{\alpha }\left( \mathbf{k},\omega \right) \omega ^{2}d\omega =2k_{\parallel
}^{2}v_{e}v_{\alpha } \int_{0}^{\infty }e^{-U_{e}\left( \mathbf{k}%
,t\right) }e^{-U_{\alpha}\left( \mathbf{k},t\right) }\left[ \frac{\partial }{%
\partial t}U_{e}\left( \mathbf{k},t\right) \right] \left[ \frac{%
\partial }{\partial t}U_{\alpha}\left( \mathbf{k},t\right) \right] dt ,
\label{eq:b1}
\end{equation}
where $U_{\alpha}(\mathbf{k},\omega)$ is given by Eq.~\eqref{eq:a16}. Substituting this expression into
Eq.~\eqref{eq:gamma2} for the Coulomb integral for the e-e collisions after $\mathbf{k}$-integration
we arrive at
\begin{equation}
\ln \Lambda _{ee}=\int_{0}^{\infty }\frac{dt}{t}\int_{0}^{1}d\mu
\frac{\phi ^{2}\left( t,\mu \right) }{\chi ^{3/2}\left( t,\mu \right) }\left[
\Phi \left( \frac{2\xi _{e} t}{\zeta _{e}}\sqrt{\chi \left( t,\mu \right) }%
\right) -\Phi \left( \frac{2t}{\zeta _{e}}\sqrt{\chi \left( t,\mu
\right) }\right) \right] .
\label{eq:b2}
\end{equation}
Here $\zeta_{e}=\lambda_{\mathrm{D}}/a_{e}$, $\xi_{e} =\lambda_{\mathrm{D}}/r_{ee ,\min}$, and
\begin{eqnarray}
&&\chi \left( t,\mu \right) =\mu ^{2}+\left( 1-\mu ^{2}\right) \left( \frac{%
\sin t}{t}\right) ^{2} ,   \label{eq:b3}    \\
&&\phi \left( t,\mu \right) =\mu ^{2}+\left( 1-\mu ^{2}\right) \frac{\sin
\left( 2t\right) }{2t} ,   \label{eq:b4}    \\
&&\Phi \left( x\right)=\frac{4}{\sqrt{\pi }}\int_{0}^{x}e^{-t^{2}}t^{2}dt =\mathrm{erf} \left( x\right)
-\frac{2}{\sqrt{\pi }} xe^{-x^{2}} , \label{eq:b5}
\end{eqnarray}
where $\mathrm{erf} (x)$ is the error function. Similarly, for the e-i collisions we find
\begin{equation}
\ln \Lambda _{ei}=\left( 1+\delta ^{2}\right) ^{3/2}\int_{0}^{\infty }\frac{dt%
}{t}\int_{0}^{1}d\mu \frac{\phi \left( t,\mu \right) \phi \left(\epsilon
t,\mu \right) }{\varphi ^{3/2}\left( t,\mu \right) }\left[ \Phi \left( \frac{%
\xi t}{\zeta_{e} }\sqrt{2\varphi \left( t,\mu \right) }\right) -\Phi \left(
\frac{t}{\zeta_{e} }\sqrt{2\varphi \left( t,\mu \right) }\right) \right]
\label{eq:b6}
\end{equation}
with $\epsilon =Z_{i}m_{e}/m_{i}$, $\delta =\sqrt{ m_{e}T_{i}/m_{i}T_{e}}$, $\zeta_{e} =\lambda_{\mathrm{D}}/a_{e}$,
$\xi =\lambda_{\mathrm{D}}/r_{ei,\min}$ and
\begin{equation}
\varphi \left( t,\mu \right) =\chi \left( t,\mu \right) +\delta ^{2}\chi
\left(\epsilon t,\mu \right) .
\label{eq:b7}
\end{equation}
It is seen that Eq.~\eqref{eq:b2} with Eqs.~\eqref{eq:b3}-\eqref{eq:b5} and Eq.~\eqref{eq:b6} with \eqref{eq:b7}
coincide with the expressions derived in Ref.~\cite{sil98} on the basis of the Landau kinetic equation.

In the case of vanishing or infinitely strong magnetic fields Eqs.~\eqref{eq:b2} and \eqref{eq:b6} become
\begin{eqnarray}
&&\ln\Lambda_{0,ee}=\ln\xi_{e},  \qquad  \ln\Lambda_{0,ei}=\ln \xi ,    \label{eq:b8}   \\
&&\ln\Lambda_{\infty ,ee}=\frac{1}{2}\ln\xi_{e}, \quad  \ln\Lambda_{\infty ,ei}=\frac{1}{2}\ln \xi , \label{eq:b9}
\end{eqnarray}
respectively. Thus, if the dynamical polarization effects are neglected, the Coulomb logarithm in the presence
of a strong magnetic field is again given by $\frac{1}{2}\ln \Lambda_{0,e\alpha }$ where, however, the quantity
$\Lambda_{0,e\alpha }$ is replaced by the usual one, i.e. $r_{\max}/r_{\min}$.

\end{document}